\documentstyle[prd,aps,
preprint,
tighten,multicol,pictex]{revtex}

\begin{document}

\draft
\preprint{\begin{tabular}{r}
KIAS-P99085 \\
hep-ph/9909429
\end{tabular}}

\title{One-loop Corrected Neutrino Masses and Mixing in 
the Supersymmetric Standard Model without R-parity }
\author{Eung Jin Chun
and Sin Kyu Kang }
\address{Korea Institute for Advanced Study \\
    207-43 Cheongryangri-dong, Dongdaemun-gu,
    Seoul 130-012, Korea \\
{\it Email addresses: ejchun@kias.re.kr, skkang@kias.re.kr}    }
\maketitle

\begin{abstract}
We provide, in the R-parity violating supersymmetric standard model, a 
comprehensive analysis for the determination of sneutrino vacuum
expectation values from the one-loop 
effective scalar potential, and also for one-loop renormalized 
neutrino masses and mixing by calculating the effective neutrino mass 
matrix in the weak basis.  Applying our results to theories with 
gauge mediated supersymmetry breaking, we show how atmospheric and 
solar neutrino oscillations can be accommodated simultaneously in 
this framework.  It is observed that the one-loop correction to
sneutrino vacuum expectation values leads to a significant
effect on the determination of  the neutrino masses and mixing.
\end{abstract}

\pacs{PACS number(s): 12.60.Jv, 11.30.Fs, 14.60.Pq}


\section{Introduction}

The minimal supersymmetric standard model (MSSM) may allow for 
explicit lepton number and thus R-parity  violation through which
neutrinos get nonzero masses and mixing \cite{HS}.  
As it is an attractive possibility to explain neutrino experiment data, 
there have been many works investigating neutrino masses from R-parity 
violation \cite{oldies}.  In recent years, further development has been 
made to analyze three neutrino masses and mixing \cite{news} accommodating
the recent observation of the atmospheric muon neutrino oscillation 
in Super-Kamiokande \cite{skam} and  other experimental data 
\cite{sola,lsnd}.  A very interesting feature of R-parity violation
as the origin of neutrino masses and mixing is that this idea could  
be tested in future collider experiments despite the small R-parity
violating couplings.  Specifically, 
the mixing angles measured in the Super-Kamiokande 
\cite{skam} and the CHOOZ experiments \cite{chooz} can be probed
by detecting lepton number violating decay modes of 
the lightest supersymmetric particle \cite{viss,jaja}.  

\medskip

In this paper, we basically extend the former studies in two ways.
First, we calculate one-loop improved tree-level neutrino masses. In
order to do that, we determine sneutrino vacuum expectation values 
(VEVs) from the minimization of the one-loop effective scalar potential,
which has usually been ignored in previous works 
\footnote{see however, Ref.~\cite{valle} where diagrammatic method is used.}.
As will be shown later, the one-loop correction to sneutrino VEVs should not be
ignored as it can drastically change the tree-level results.
When there exists a cancellation among tree-level quantities
in the determination of sneutrino VEVs, the loop contributions to the neutrino
masses can be even larger than the tree ones.  
Furthermore, this phenomenon occurs generically in a certain parameter region.  
Secondly, we calculate one-loop renormalized neutrino masses and mixing
utilizing the ``effective mixing matrix'' method \cite{nojiri}.
As far as one concerns only the neutrino sector, this approach turns out
to be very useful even though the results should be the same as in the usual
on-shell renormalization scheme working in the tree-level mass basis
\cite{hemp,valle}.
Our procedure begins with constructing the 7x7 one-loop neutrino/neutralino 
mass matrix in the weak basis. Then, we obtain the 
seesaw-reduced 3x3 neutrino mass matrix from which physical neutrino 
masses and mixing are calculated.  In this way, one can easily figure out 
the effects of each self-energy diagrams and simplify the calculation to
obtain the main contributions.
Our results are applied to the theories with gauge mediated supersymmetry
breaking to discuss how large the one-loop results deviate from the  
tree-level ones.  Finally, we show that the atmospheric and solar
neutrino oscillations can be accommodated naturally in this framework.
In particular, we present some typical examples for  input parameters 
providing the vacuum or matter oscillation solution to the solar 
neutrino problem.

\medskip

This paper is organized as follows.  In Section II, we determine
sneutrino VEVs by computing one-loop effective scalar potential, and 
present analytic expressions for all the one-loop contributions 
to sneutrino VEVs in Appendix A.
In doing so, we also derive the diagonalization matrices for fermions or
sfermions which get mixed further due to R-parity violation.  In Section III,
these diagonalization matrices will be used to  calculate 
one-loop self-energy diagrams in the weak basis.  
This enables us to determine the dominant contributions to physical 
neutrino mass eigenvalues and mixing angles which could be  relevant 
for explaining the current neutrino data.
The leading self-energy diagrams are calculated  in Appendix B.
In Section IV, we apply the above results to the R-parity violating
theories with gauge mediated supersymmetry breaking and discuss 
the effects of one-loop corrections to sneutrino VEVs on the
determination of the  neutrino masses and mixing.
Then, we illustrate how the atmospheric and solar neutrino data 
can be explained simultaneously in this framework. 
We conclude in Section V.

\section{Sneutrino VEVs from one-loop effective scalar potential}

We start by writing the MSSM superpotential $W_0$ and $W_1$ conserving
and violating R-parity/lepton number, respectively:
\begin{eqnarray} \label{supo}
W_0 &= & \mu H_1 H_2 + h^e_i L_i H_1 E^c_i+ h^d_i Q_iH_1 D^c_i 
       + h^u_i Q_iH_2 U^c_i               \nonumber\\
W_1  &= & \mu_i L_i H_2 + \lambda_{ijk} L_i L_j E^c_k 
     + \lambda'_{ijk} L_i Q_j D^c_k  \,.
\end{eqnarray}
An unpleasant feature of R-parity violation is that there are too many free
parameters essentially uncontrollable unless a certain theoretical framework
for Yukawa hierarchies is invoked.
For later use, we assume that  $\lambda_i\equiv \lambda_{i33}$ and
$\lambda'_i\equiv \lambda'_{i33}$ dominate over the other couplings
for the first two generations, guided by the quark and lepton Yukawa
hierarchy.  The bilinear model where no trilinear R-parity violating 
couplings are present at the fundamental scale satisfies 
this assumption,  as rotating away the bilinear terms  generates
the hierarchical trilinear couplings $\lambda,\lambda'$.

One-loop effective scalar potential to determine the VEVs of the Higgs and
sneutrino fields, $\langle H_{1,2}^0 \rangle$ and $\langle L_i^0 \rangle$ 
is given by
\begin{equation} \label{Veff}
 V= [ m^2_{L_i H_1} L_i H_1^\dagger + B_i L_i H_2 + {\rm h.c.} ] + m^2_{L_i} 
    |L_i|^2 + V_D + V_1 + \cdots \,,
\end{equation}
where $V_D$ is the $SU(2)\times U(1)$ D-term, and $V_1={1\over 64\pi^2} 
{\rm Str}{\cal M}^4\left(\ln {{\cal M}^2\over Q^2} -{3\over2}\right)$ is the 
one-loop correction.
In our notation, $m^2_{L_i H_1}$ contains not only the soft mass term
 but also the supersymmetric mass term $\mu\mu_i$ from the superpotential
(\ref{supo}).
Here we assume the R-parity violating parameters are small enough,
$\mu_i \ll \mu$ and $m^2_{L_iH_1} \ll m^2_{L_i}$, etc., to produce
light neutrinos of $m_{\nu} <1$eV, which would be a consequence of
a horizontal symmetry \cite{flasy}.
Furthermore, we take the universality among soft supersymmetry breaking
terms at the scale of supersymmetry breaking where the bilinear term
$\mu_i L_i H_2$ can be rotated away together with the corresponding
soft terms.
One of a natural framework for the universality is gauge mediated
supersymmetry breaking which will be discussed in Section IV.
Under these assumptions the intergenerational mixing terms like
$\mu_i \mu_i L_i L_j^{\dagger}$ can be neglected.
The minimization condition for the sneutrino fields,
$\partial V/\partial L^0_i=0$, leads to
\begin{equation} \label{svev}
 \langle L_i^0 \rangle = 
 -{ B_i \langle H_2^0 \rangle + m^2_{L_i H_1} \langle H_1^0 \rangle 
          + \Sigma^{(1)}_{L_i} \langle H_1^0 \rangle 
       \over m^2_{L_i} + {1\over2} M_Z^2 c_{2\beta} + \Sigma^{(2)}_{L_i} }\,,
\end{equation}
where  $\Sigma^{(1,2)}_{L_i}$ are given by
\begin{equation} \label{Sigme}
 \Sigma^{(1)}_{L_i} = 
   \left.{\partial V_1 \over H_1^{0*} \partial L_i^0}\right|_0, \quad
 \Sigma^{(2)}_{L_i} = 
   \left.{\partial V_1 \over L_1^{0*} \partial L_i^0}\right|_0  \,,
\end{equation}
where $\left. \right|_0$ means to put $L_i^0$ to zero. 
This expression is valid when R-parity violating parameters are small enough, 
$\mu_i \ll \mu$, etc..
In the case of small R-parity violation, the Higgs VEVs can be obtained from
the usual minimization condition neglecting  R-parity violating couplings. 
The complete one-loop calculation for the Higgs VEVs
can be found in Ref.~\cite{hvev}.

R-parity violation mixes  neutrinos, charged leptons, sleptons 
and sneutrinos with neutralinos, charginos, charged and neutral Higgses, 
respectively, and gives rise to  more complicated and
bigger mass matrices of fermions and sfermions.  
In the case of small R-parity violation, 
it is convenient to use first approximate seesaw diagonalization 
to rotate away the R-parity odd elements.  Then one uses the further
diagonalization process to obtain mass eigenvalues which contain the 
contributions from R-parity violating parameters up to their second order.  
It is now a simple manner to take derivatives of mass eigenvalues 
with respect to $L_i^0$ and find out the quantities in Eq.~(\ref{Sigme}).

\medskip

Let us follow the above prescription first to get charged slepton/Higgs
contribution to sneutrino VEVs.  Since sleptons and charged Higgs bosons
get mixed, one has to diagonalize the $8\times8$ mass matrix for 
three slepton pairs and one Higgs pair which takes  the form:
\begin{equation} \label{MlH}
 {\cal M}^2=\pmatrix{ {\cal M}^2_{\tilde{l}\tilde{l}} & 
                      {\cal M}^2_{\tilde{l}H} \cr
		      {\cal M}^{2\dagger}_{\tilde{l}H} &
		      {\cal M}^2_{HH} \cr } \,.
\end{equation}
To calculate $\Sigma^{(1,2)}$ we keep the R-parity violating contributions
to the above mass matrix up to quadratic terms in R-parity violating
parameters.    In this paper, we assume the absence of intergenerational 
mixing in slepton soft masses.  However, there could arise in general 
R-parity violating intergenerational mixing masses
which can be neglected as they give quartic contribution to mass eigenvalues.
We further assume CP conservation and thus take the VEVs,
$u_i=\langle L_i^0 \rangle$, $v_{1,2}=\langle H_{1,2}^0 \rangle$ to be real.  
Then,  the submatrices of Eq.~(\ref{MlH}) in the weak basis 
$(\tilde{e}_{iL}, \tilde{e}_{iR}, H_1^-, H_2^{-})$ are given by
\begin{eqnarray} \label{MlHs}
&& {\cal M}^2_{\tilde{l}\tilde{l}} =
 \pmatrix{ M^2_{Li} -(g_w^2+{1\over2}g_z^2) u^2 + g_w^2 u_i^2 
           -2  h^e_i v_1 ( \lambda_j u_j) &
            M^2_{Di} + \delta_{i3} ( A_j u_j)       \cr
            M^2_{Di} + \delta_{i3} ( A_j u_j)    & 
	    M^2_{Ri} +(g_w^2-g_z^2) u^2 + ( h^e_i u_i)^2
	    - 2  h^e_i v_1 ( \lambda_j u_j) \cr } \nonumber\\
&& {\cal M}^2_{\tilde{l}H} =
  \pmatrix{ m^2_{L_i H_1} +(g_w^2-h^{e2}_i) v_1 u_i &
            -B_i+g_w^2 u_i v_2  \cr
	    h^e_i v_2 \mu_i + A^e_i u_i & 
	    h^e_i v_1 \mu_i-h^e_i \mu u_i \cr } \\
&& {\cal M}^2_{HH} =
   \pmatrix{ M^2_{H_1^-} -{1\over2}(2g_w^2-g_z^2) u^2 +h_\tau^2 u_3^2 &
            -B + g_w^2 v_1 v_2  \cr -B + g_w^2 v_1 v_2 & 
	    M^2_{H_2^+} +{1\over2}(2g_w^2-g_z^2)u^2 \cr } \nonumber \,,
\end{eqnarray}
where $M^2_{Li}$, {\it etc} are the conventional R-parity conserving
charged slepton or Higgs masses and summing over $j$ is understood
for the quantities inside the parentheses in the first matrix, 
and we will take the lepton Yukawa couplings, $h^e_{1,2}=0$.  
Here we define the notation $g_w^2 \equiv M_W^2/v^2$
and $g_z^2\equiv M_Z^2/v^2$ where $v^2=v_1^2+v_2^2+u^2$ and
$u^2=\sum_i u_i^2$.  Following the previous prescription,
one can derive the charged slepton and Higgs contributions
to $\Sigma^{(1,2)}$ as presented in Appendix A.
Here we have to mention a few technical points.
For diagonalizing away the R-parity odd matrix elements, we use the seesaw 
formula.  That is, the matrix ${\cal M}$ of the form;
\begin{equation}
 {\cal M} = \pmatrix{ M_1 & \Delta \cr \Delta^\dagger & M_2 }
\end{equation}
can be diagonalized by the ``approximate'' unitary matrix ${\cal U}$;
\begin{equation} \label{Thet}
 {\cal U} = \pmatrix{ 1- {1\over2} \Theta \Theta^\dagger & -\Theta \cr
            \Theta^\dagger & 1-{1\over2} \Theta^\dagger \Theta \cr} 
\end{equation}
if $M_{1,2}$ and $\Delta$ are submatrices satisfying $\Delta \ll M_{1,2}$.
Being valid up to quadratic order in $\Delta$, the seesaw rotation matrix
$\Theta$ is determined by the relation, $\Delta=M_1 \Theta-\Theta M_2$ and 
the block diagonal matrices become
\begin{eqnarray} \label{M12}
 M_1 &\to& M_1 +          \nonumber
      {1\over2}\left(\Theta \Delta^\dagger+\Delta \Theta^\dagger\right)  \\
 M_2 &\to& M_2 - 
      {1\over2}\left(\Theta^\dagger \Delta+\Delta^\dagger \Theta\right)  \,.
\end{eqnarray}
The essential step in the diagonalization 
is to decouple the Goldstone mode, $G^-$.
This can be done by first going to the basis of vanishing sneutrino VEVs
with the rotation matrix $\Theta_G$ and then applying the usual rotation
$U_\beta$ with the angle $\beta$.  Omitting the quadratic terms in R-parity
violating parameters for simplicity, the combined rotation 
$\Theta_G U_\beta$  is given by
\begin{equation} \label{GlH}
 \Theta_G U_\beta= \pmatrix{1 &0 & a_i &0 \cr
            0 & 1 & 0 & 0 \cr -a_i & 0 & 1 & 0 \cr
	    0 & 0 & 0 & 1 \cr } 
	    \pmatrix{ 1 & 0 & 0 & 0 \cr 0 & 1 & 0 & 0 \cr
	             0 & 0 & -c_\beta & s_\beta \cr
		     0 & 0 & s_\beta  & c_\beta \cr } \,,
\end{equation}
where $a_i \equiv u_i/v_1$ are the values determined from tree level
scalar potential.
Note that in the full expression, the angle $\beta$ is determined by
$\tan\beta=v_2/v'_1$ where $v'_1=\sqrt{v_1^2+u^2}$.
Applying the above rotation to the charged lepton/Higgs mass matrix 
(\ref{MlH},\ref{MlHs}), one can explicitly see that there is a massless
state up to the order of our expansion.
Further diagonalization can be done by another seesaw rotation
$\Theta_\delta$ which get rid of R-parity odd off-diagonal components,
and the usual diagonalization $U_{e_i}$ of the  2x2 slepton mass matrix 
[see Appendix A].  Again in the leading approximation,
$\Theta_\delta U_{e_i}$ is given by  
\begin{equation}
 \Theta_\delta U_{e_i} =
 \pmatrix{ 1 & 0 & 0 & -\delta_1^i \cr 0 & 1 & 0 & -\delta_2^i \cr
           0 & 0 & 1 & 0 \cr \delta_1^i & \delta_2^i & 0 & 1 \cr }
 \pmatrix{ c_{e_i}  &  s_{e_i}  & 0 & 0 \cr -s_{e_i} & c_{e_i}  & 0 & 0 \cr
           0 & 0 & 1 & 0 \cr 0 & 0 & 0 & 1 \cr }
\end{equation}
We obtain the following expressions for $\delta_{1,2}^i$;
\begin{eqnarray} \label{deltas}
 M^2_{E_iH}= -a_i s_\beta(m^2_{H^-}-M^2_W)-{B_i \over c_\beta}, &\quad&
  M^2_{E^c_iH}= (a_i- {\mu_i\over \mu}) {m^e_i \mu \over c_\beta}   
           \nonumber\\
 \delta_1^i = {M^2_{E_iH}(M^2_{Ri}-m^2_{H^-}) - M^2_{E^c_i H} M^2_{Di} \over
     (m^2_{\tilde{e}_{i1}}-m^2_{H^-}) (m^2_{\tilde{e}_{i2}}-m^2_{H^-}) } \,,
    &\quad&
 \delta_2^i = {M^2_{E^c_iH}(M^2_{Li}-m^2_{H^-}) - M^2_{E_i H} M^2_{Di} \over
     (m^2_{\tilde{e}_{i1}}-m^2_{H^-}) (m^2_{\tilde{e}_{i2}}-m^2_{H^-}) }  
\end{eqnarray}
In sum, the rotation matrix which brings the weak eigenstates
($\tilde{e}_{iL}, \tilde{e}_{iR}, H_1^-, H_2^{-}$) to the mass eigenstates
($\tilde{e}_{i1}, \tilde{e}_{i2}, H^-$) decoupling  the Goldstone mode $G^-$
is 
\begin{equation} \label{UC}
 {\cal U}^C =
 \pmatrix{ c_{e_i} & s_{e_i} &  -\delta_1^i + a_i s_\beta \cr
           -s_{e_i} & c_{e_i} &  -\delta_2^i \cr
      (-a_i +  \delta_1^i s_\beta)c_{e_i} -\delta_2^i s_\beta s_{e_i}&
     (-a_i +  \delta_1^i s_\beta) s_{e_i}+\delta_2^i s_\beta c_{e_i} &
      s_\beta \cr \delta_1^i c_\beta c_{e_i}-\delta_2^i  c_\beta s_{e_i}&
     \delta_1^i c_\beta s_{e_i}+\delta_2^i c_\beta c_{e_i}& 
	   c_\beta \cr} \,.
\end{equation}
This matrix will be used to compute one-loop self-energy diagrams for
neutrino/neutralino masses.

Following the similar steps, we also obtain the CP-even and CP-odd 
sneutrino/Higgs contributions to sneutrino VEVs.
The mass matrix in the weak basis $(\tilde{\nu}_i^R, H_1^R, H_2^R)$ of
the real components of slepton and Higgs fields is
\begin{equation} \label{MRH}
 {\cal M}_{\tilde{\nu}^RH} = 
 \pmatrix{ M^2_{\tilde{\nu}_i} + {g_z^2\over2}u^2+g_z^2u_i^2 &
           m^2_{L_iH_1}+g_z^2v_1u_i & B_i-g_z^2u_iv_2 \cr
	    m^2_{L_iH_1}+g_z^2v_1u_i & M^2_{H_1^R}+{g_z^2\over2} u^2 &
	   B-g_z^2v_1v_2 \cr B_i-g_z^2u_iv_2 &  
	   B-g_z^2v_1v_2 &  M^2_{H_2^R}-{g_z^2\over2} u^2 \cr }
\end{equation}
where $M^2_{\tilde{\nu}_i}$, {\it etc} are the usual R-parity conserving
sneutrino/neutral Higgs  boson masses. The diagonalization matrix in the leading
expansion in terms of R-parity violating parameters is
\begin{equation} \label{US}
 {\cal U}^S =
 \pmatrix{ 1 & \zeta_1^i s_\alpha-\zeta_2^i c_\alpha & 
           -\zeta_1^i c_\alpha - \zeta_2^i s_\alpha \cr
	   \zeta_1^i & -s_\alpha & c_\alpha \cr
	   \zeta_2^i & c_\alpha & s_\alpha \cr }
\end{equation}
where $\alpha$ is the usual neutral Higgs scalar diagonalization angle 
[see Appendix A] and the seesaw rotation angle $\zeta_{1,2}^i$ is given 
by 
\begin{eqnarray} \label{zeta12}
 &&\zeta_1^i= -a_i+ { -a_i( m^2_{\tilde{\nu}_i}+M_Z^2)m_A^2s_\beta^2
                    -B_i t_\beta(m^2_{\tilde{\nu}_i}+M^2_Zc_{2\beta}) \over
		    m^4_{\tilde{\nu}_i}-m^2_{\tilde{\nu}_i}(m^2_{h}+m^2_H)
		    +m^2_h m^2_H } \nonumber\\
 &&\zeta_2^i=  { a_i( m^2_{\tilde{\nu}_i}-M_Z^2 c_{2\beta} )m_A^2s_{2\beta}/2
                 +B_i (m^2_{\tilde{\nu}_i}-M^2_Zc_{2\beta}) \over
		    m^4_{\tilde{\nu}_i}-m^2_{\tilde{\nu}_i}(m^2_{h}+m^2_H)
		    +m^2_h m^2_H } 
\end{eqnarray}

As for the  CP-odd sneutrino/Higgs fields,
the mass matrix in the weak basis $(\tilde{\nu}^I, H_1^I, H_2^I)$ of
the imaginary components of slepton and Higgs fields is
\begin{equation} \label{MIH}
 {\cal M}_{\tilde{\nu}^IH} = 
 \pmatrix{ M^2_{\tilde{\nu}_i} + {g_z^2\over2}u^2 &
           m^2_{L_iH_1} & -B_i \cr
	    m^2_{L_iH_1} & M^2_{H_1^R}+{g_z^2\over2} u^2 &
	   -B \cr -B_i &  
	  -B &  M^2_{H_2^I}-{g_z^2\over2} u^2 \cr }
\end{equation}
where $M^2_{\tilde{\nu}^I_i}$, {\it etc} are the usual R-parity conserving
sneutrino/neutral Higgs masses. As in the charged slepton/Higgs case,
we first make the seesaw rotation, composed of $a_i=u_i/v_1$
to decouple the Goldstone mode, $G^0$.
The diagonalization matrix 
which brings the weak eigenstates to the mass 
eigenstates $(\tilde{\nu}^I, A^0)$ becomes
\begin{equation} \label{UP}
 {\cal U}^P =
 \pmatrix{ 1 &   -\zeta_3^i + a_i s_\beta \cr
	   -a_i + \zeta_3^i s_\beta &  s_\beta  \cr
	   \zeta_3^i  c_\beta & c_\beta \cr } \,,
\end{equation}
where the seesaw rotation angle $\zeta_{3}^i$ is given by
\begin{equation} \label{zeta3}
 \zeta_3^i = {B_i-a_i B \over c_\beta (m_A^2-m^2_{\tilde{\nu}_i}) } \,.
\end{equation}

In Appendix A, we collect all of the information needed to calculate one-loop
corrected sneutrino VEVs and present complete analytic results.
For the numerical calculation, we take the renormalization scale,
$Q^2=m_{\tilde{t}_1} m_{\tilde{t}_2}$, around which the one-loop 
contribution $V_1$ is minimized.
The mixing matrices given in Eqs.~(\ref{UC}), (\ref{US}) and (\ref{UP}) 
play a role of inducing  the effective $\lambda$ or $\lambda'$ couplings 
in the mass basis and will be used to calculate the one-loop 
neutrino masses.  We also need the neutrino/neutralino and 
charged lepton/chargino mixing matrices \cite{NP,jaja}
which we summarize in Appendix A for completeness.

\section{One-loop corrected neutrino mass matrix}

The neutrino/neutralino mass matrix at one-loop level is given by
\begin{equation} \label{Mpole}
 M^{pole}(p^2) = M(Q)+ \Pi(p^2) - 
          {1\over2}\left( M(Q) \Sigma(p^2)+ \Sigma(p^2) M(Q)\right)
\end{equation}
where  $Q$ is the renormalization scale, and $M(Q)$ is the
the $\overline{DR}$ renormalized 7x7 tree-level mass matrix, $\Pi$ and $\Sigma$
are the contributions from one-loop self-energy diagrams.
To compute the one-loop corrected mass matrix (\ref{Mpole}), we work in the 
weak basis, that is, we evaluate $\Pi$'s and $\Sigma$'s with
weak eigenstate neutrino/neutralino on external legs and mass eigenstates 
running inside loops. For this, we express the propagators in the
loops in terms of the mass eigenstates as calculated in Appendix B.
A complete one-loop analysis has been performed in Refs.~\cite{hemp,valle}
where the mass matrix (\ref{Mpole}) is calculated in the tree-level mass basis
and mass eigenvalues and mixing matrix are obtained by re-diagonalizing the
one-loop mass matrix.
In our approach, it is useful to make a few observations which facilitate
out calculation.
Let us recast the mass matrix (\ref{Mpole}) as
\begin{eqnarray} \label{Mpole2}
 M^{pole}(p^2) &=&  \pmatrix{ 0 & M_D \cr M_D^T & M_N \cr }(Q)
    + \pmatrix{ \Pi_n & \Pi_D \cr \Pi_D^T & \Pi_N \cr }(p^2)  \nonumber\\
&&    - \frac{1}{2}\pmatrix{ M_D \Sigma_D^T+\Sigma_D M_D^T & 
                \Sigma_n M_D+M_D \Sigma_N + \Sigma_D M_N \cr 
               M_D^T \Sigma_n + \Sigma_N M_D^T+ M_N\Sigma_D^T  & 
                M_N \Sigma_N+\Sigma_N M_N  \cr }(p^2)
\end{eqnarray}
where neutrino--neutralino mixing elements of $M_D$ or $\Pi_D, \Sigma_D$ are
much smaller than $M_N$ or $\Pi_N, \Sigma_N$ by factor of ${\cal O}(a_i)$.
If we are interested only in neutrino mass eigenvalues and mixing matrix, 
it is convenient to obtain first the seesaw reduced 3x3 neutrino mass matrix.
Applying seesaw formula to Eq.~(\ref{Mpole2}) and keeping one-loop order
terms, we find the effective 3x3 neutrino mass matrix as follows;
\begin{eqnarray} \label{mpole}
M^{\nu} (p^2) &=&  -M_D M^{-1}_N M_D^T(Q)
   + M_D M_N^{-1} \Pi_N\!(p^2) M^{-1}_N M_D^T  \nonumber\\
 &&  +(M_D M^{-1}_N M_D^T \Sigma_n\!(p^2)
       + \Sigma_n\!(p^2) M_D M^{-1}_N M_D^T) \\
&&  +(M_D M^{-1}_N \Pi_D^T\!(p^2)+ \Pi_D\!(p^2) M^{-1}_N M_D^T)
    + \Pi_n\!(p^2) \nonumber
\end{eqnarray}
As neutrino masses are negligible, we can now take $p^2=0$ without ambiguity
to calculate physical neutrino masses and mixing.  Our procedure is in fact 
identical to that developed in Ref.~\cite{nojiri} where the mixing 
renormalization is performed by utilizing the $Q$-independent ``effective
mixing matrix''.  The effective mixing matrix $U^{eff}$ is calculated from 
the diagonalization matrix $U(p^2)$ of the mass matrix $M^{pole}(p^2)$ in 
the weak basis (\ref{Mpole}) by taking $U^{eff}_{ij}=U_{ij}(m_j^2)$.
According to the results of \cite{nojiri},  this method has several nice
features. First, one can avoid superficial
singularity reflecting the arbitrariness in the diagonalization of
the mass matrix with degenerate massless neutrinos at tree level.
Second, as far as neutrino mixing is concerned, the mixing matrix
$U^\nu=U(p^2\to0)$ determined from the diagonalization of the effective
3x3 matrix (\ref{mpole}) can be considered as the full unitary mixing matrix,
up to a negligible correction of ${\cal O}(a_i^2)$.  Furthermore, the
scale dependence which appears only in $\Sigma_n$ turns out to be very
small as we will discuss below.

Note that the first term in Eq.~(\ref{mpole}) is the tree-level mass 
which makes only one neutrino massive.
The other terms show one-loop contributions from various self-energy diagrams.  
It is interesting to see that $\Sigma_{D,N}$ do not appear in $M^\nu(p^2)$
and thus one does not have to calculate them.
We can make further simplifications.
First, in the basis where $\mu_i$ terms
are rotated away, not only the tree-level mass matrix 
but also the loop mass matrix in Eq.~(\ref{mpole})
take a factorized form $M^{\nu}_{ij} \propto a_i a_j$ \cite{jaja}, 
That is, the contribution from $\Pi_N$ is aligned with the tree-level 
mass matrix and thus gives overall correction to the dominant tree-level mass 
in the direction $ \sim a_i \nu_i$.
One-loop correction to neutralino masses has been analyzed in 
Ref.~\cite{pierce}, according to which the $\Pi_N$ contribution 
typically gives 6\% correction to the tree-level values.
As we are interested in the leading correction to the masses of the two 
light neutrinos, we will ignore this correction by working 
in the basis with $\mu_i=0$.  
Secondly,  once no flavor changing sfermion masses are assumed, 
$\Sigma_n$ has only diagonal elements
which produces also aligned correction of the order $\alpha_2/4\pi
\lesssim1\%$, and thus we can also ignore this contribution.
Then we are left with the last two one-loop corrections involving
$\Pi_n$ and $\Pi_D$ in Eq.~(\ref{mpole}) which will give rise 
to a sizable contribution to neutrino masses and mixing.
Therefore,  the resulting neutrino mass matrix is  
\begin{eqnarray} \label{mn}
 M^{\nu}_{ij} &=& {M^2_Z \over F_N} \xi_i \xi_j c_\beta^2
            + {M^2_Z \over F_N} (\xi_i \eta_j +\eta_i \xi_j) c_\beta 
                 + \Pi_{\nu_i \nu_j}\quad{\rm where}
		 \nonumber \\
\quad F_N &=&-( M_1M_2/M_{\tilde{\gamma}}+M_Z^2s_{2\beta}/\mu), \quad
                M_{\tilde{\gamma}}= M_1 c_W^2+ M_2 s_W^2 \\
 \eta_i&=&\Pi_{\nu_i \tilde{B}^0}({-M_2 s_W^2 \over M_{\tilde{\gamma}}M_Wt_W})
 +\Pi_{\nu_i \tilde{W}_3} ({M_1 c_W^2 \over M_{\tilde{\gamma}} M_W })
 +\Pi_{\nu_i \tilde{H}_1^0} ({s_\beta \over \mu})
 +\Pi_{\nu_i \tilde{H}_2^0} ({-c_\beta \over \mu})  \nonumber
\end{eqnarray}
Note that $\xi_i\equiv a_i-\mu_i/\mu$ becomes $a_i$ in the basis with
$\mu_i=0$.
Let us now make a qualitative discussion on various contributions to
$\eta_i$ and $\Pi_{\nu_i \nu_j}$.

\medskip

The one-loop diagrams contributing to $\Pi_D$ (or $\Pi_{\nu \psi^0}$)
can be divided into
[DD], [DF] and [FF] types which involve two gauge vertices, one gauge and
one Yukawa vertices, and two Yukawa vertices \cite{HS}.  The explicit formulae
for $\Pi_D$ (and also $\Pi_n$) are presented in Appendix B.
The [DD] type contains diagrams with $Z/W$, sneutrino/Higgs, charged slepton/
Higgs exchanges.  Among them, the Z/W exchanging diagrams  give corrections of
${\cal O}({\alpha_2\over4\pi}{m_{\tau}^2 t_\beta \over M_W^2}) \lesssim
10^{-4}$ which are aligned with the tree level masses and thus
can be safely ignored. The contribution from  the charged slepton/Higgs
exchange diagrams (involving also charged lepton/chargino exchange)
can be as large as ${\alpha_2 \over 4\pi} t_\beta$ apart from the possible 
alignment as can be read from mixing matrices (\ref{deltas},\ref{UC}) and 
(\ref{ULR},\ref{rls}).  The same is true for the sneutrino/Higgs
exchanging diagrams (involving also neutrino/neutralino exchange).
 From our explicit calculation in the next section, we will see that
the seemingly large correction of order ${\alpha_2 \over 4\pi} t_\beta$ 
for large $\tan\beta$ is compensated by a large suppression  factor 
and thus $\xi_i \gg \eta_i$.
The [DF] type includes the diagrams with $h_\tau$ coupling involving
charged slepton/Higgs exchange, $\lambda$ coupling involving
stau/tau exchange, and $\lambda'$ coupling involving sbottom/bottom
exchange. The [DF] diagrams with $h_\tau$ coupling can be negligible for small
$\tan\beta$ as the correction is of the order ${gh_\tau \over 16\pi^2} 
{m_\tau \over M_W} \lesssim 10^{-5}$ and are sub-leading to
the corresponding diagrams contributing to  $\Pi_n$ for large $\tan\beta$.
In the [FF] type, there are  diagrams with $h_\tau h_\tau$, 
$h_\tau \lambda$ and $h_b \lambda'$ couplings.  
Generically, these diagrams are sub-leading to the similar ones in $\Pi_n$.

There  are also [DD], [DF], and [FF] types in the $\Pi_n$ (or $\Pi_{\nu\nu}$)
diagrams.  The [DD] type can be ignored as 
the Z exchanging diagrams give aligned corrections of the order of 1 \%  and
the sneutrino exchanging ones are also negligible.
In the [DF] type, there are  diagrams with the coupling $h_\tau$ and 
$\lambda$. The first ones give rise to the correction $\sim {gh_\tau
\over 16\pi^2}{m_\tau t_\beta \over  M_W}$ which has $\tan\beta$ 
enhancement compared to the corresponding diagrams in $\Pi_D$ as mentioned
before.  In the [FF] type, there are  diagrams involving $h_\tau h_\tau$,
$h_\tau \lambda$, $\lambda\lambda$ and $\lambda'\lambda'$ couplings.
The first two classes of diagrams may have significant contributions
for large $\tan\beta$ as noted in Refs.~\cite{hwang}, which will be
partly confirmed in the next section.  
The last two classes of diagrams are the frequently considered ones as
they typically give the main contributions to the one-loop correction.

Our qualitative discussion on the one-loop corrected neutrino masses
will be confirmed by numerical computation performed  in the next section.
For our explicit computation, we use gauge-mediated supersymmetry
breaking models \cite{mgm} where the universality arises more naturally
than in supergravity models.
A rigorous calculation in minimal supergravity models with bilinear terms
has been performed in Ref.~\cite{valle}.

\section{Atmospheric and Solar neutrino oscillations in 
 gauge-mediated supersymmetry breaking models}

In this section, we discuss the properties of the one-loop improved 
neutrino mass matrix obtained  in the previous section through 
explicit calculation. Then, our emphasis is put on addressing 
whether or how the atmospheric and solar neutrino masses and mixing  angles
can be realized  in the context of R-parity violating models 
with supersymmetry broken by the gauge mediation mechanism.  
In a gauge mediation model, soft parameters of sfermion fields depend
only on their gauge quantum numbers upon supersymmetry breaking,
and thus the supersymmetric flavor problem is resolved 
in a natural way \cite{gmsb}.  
However, a difficulty in this type of models is that 
the origin of $\mu$ and $B$ terms cannot be provided in its own context.
This also applies to the R-parity violating bilinear terms $\mu_i, B_i$.
Assuming, however, $B$ and $B_i$ terms come from the same origin, we take
$B/B_i=\mu/\mu_i$.  Then, one can rotate away the R-parity 
violating bilinear terms into the Higgs bilinear term at the mediation scale
and thus no sneutrino VEVs can be generated.  But this property 
is spoiled by renormalization group evolution and thus sneutrinos get nonzero
VEVs at the weak scale.  We solve renormalization group equation 
to determine these sneutrino VEVs for given input parameters $\mu_i$ or
$\lambda', \lambda$.   In the R-parity preserving case,  
an extensive study of sparticle spectrum through
renormalization group evolution has been done in Ref.~\cite{wells}.
In this work, we take a minimal gauge mediation model assuming
one pair of messenger multiplets $({\bf 5}+\bar{\bf 5})$ \cite{mgm}
and put the messenger scale equal to the messenger sector supersymmetry
breaking scale $\Lambda$.  In most cases, we will set $\Lambda=75$ TeV.

The simplest R-parity violating model is the ``bilinear model'' 
in which only bilinear R-parity violating terms are present
at the fundamental scale, or at the mediation scale of supersymmetry 
breaking in our case.  This model having only three R-parity violating
parameters is more predictive and turns out to be very restrictive 
as we will show.
It is also conceivable that there are both bilinear and trilinear
couplings originated from a certain flavor symmetry which also explains 
quark and lepton mass hierarchies \cite{flasy}.  
In this spirit, we assume that $\lambda_i=\lambda_{i33}$
and  $\lambda'_i=\lambda'_{i33}$  are much larger than the other 
couplings and give rise to experimentally relevant contributions.
We call this the ``trilinear model'' which  has five R-parity violating
parameters (note that $\lambda_3\equiv0$), as the bilinear
terms $\mu_i$ can be rotated away into the trilinear couplings.
For our purpose, it is then convenient to use renormalization group equations
in the basis where the bilinear terms $\mu_i$ are rotated away at all
the scales \cite{cw}, which are recollected in Appendix C.

\medskip

Before investigating realistic solutions to neutrino problems, it is worth 
discussing how much one-loop corrections can change tree-level results.
As an example, we take a bilinear model with $\mu_1=\mu_2=\mu_3=10^{-4}\mu$ 
at the mediation scale $\Lambda=75$ TeV.  By rotating away $\mu_i$, we
have
\begin{equation}
  \lambda'_i = {\mu_i \over \mu} h_b\,, \quad
  \lambda_i =  {\mu_i \over \mu} h_\tau  \,.
\end{equation}
In Table I, we show the values of R-parity violating parameters 
and one-loop contributions to sneutrino VEVs calculated at
$Q^2=m_{\tilde{t}_1} m_{\tilde{t}_2}$. In the table, $\tilde{q}$, $\phi^-$,
$S^0$, $P^0$, $\psi^0$, $\psi^-$, and $Z/W$ denote the
squark, charged slepton/Higgs, CP-even and CP-odd  sneutrino/Higgs, 
neutrino/neutralino, charged lepton/chargino and gauge boson contributions
to $\Sigma_{L_i}$, respectively.  The main contribution to $\Sigma^{(1)}$
comes from  stops or sbottoms, and we do not show vanishing or negligible
contributions from $\psi^0$ or bottom quark, {\it etc}.  For small $\tan\beta$,
$\Sigma^{(1)}$ gives about 18\% correction to the tree-level quantity,
$B_it_\beta+m_{L_iH_1}$ [see Eq.~(\ref{svev})].  The dominant contribution
to $\Sigma^{(2)}$ comes from charged slepton/Higgs.  Contrary to the case
of  $\Sigma^{(1)}$, the squark contribution is negligible as there 
is an almost precise cancellation between up-type and down-type squark
contributions.
The correction $\Sigma^{(2)}$ to $m^2_{\tilde{\nu}}$ is typically small
and 5\% in our case.  Taking both one-loop contributions $\Sigma^{(1,2)}$ 
into account, the one-loop improved sneutrino VEVs, $\xi_i$, get increased
by 13 \% from the tree-level value $\xi^0_i$ for $\tan\beta=3$. 
This corresponds to a 27\% increase of the tree-level neutrino mass
as it is proportional to $\xi_i^2$.  This is a significant change
when one tries to explain the experimental data.   More seriously,
the one-loop correction could even upset the tree-level calculation.
As can be seen from Table I with $\tan\beta=30$, the one-loop improved 
value $\xi_i$ is about 7 times larger than the tree-level value 
$\xi^0_i$.  This phenomenon occurs due to a cancellation between 
the tree-level quantities.  As one can see from Table I, even though 
each quantity $m^2_{L_iH_1}$ or $B_i t_\beta$ is about 10 times larger 
than the one-loop correction $\Sigma^{(1)}$,  there exists
a cancellation of first two digits in the sum, $B_it_\beta+m_{L_iH_1}$.
Therefore, it is essential to include one-loop corrections in the
calculation of neutrino masses and mixing arising from R-parity violation.

To show the $\tan\beta$ dependence of tree and loop masses,
we present order of magnitudes of some mass matrix components 
in the last five rows of Table I.  Let us first note that 
while $\eta_i$ get increased mildly with $\tan\beta$, 
the components of $\Pi_n$ with $h_\tau\lambda_i$ couplings get 
increased tremendously with $\tan\beta$ and become comparable to
the contributions from $\lambda\lambda$ couplings, as was first noted
in Ref.~\cite{hwang}.
While the heaviest neutrino mass dominated by the tree-level value 
gets little changed, the second largest neutrino mass strongly depends 
on $\tan\beta$.

\medskip

Let us now turn to an important question whether a realistic
neutrino mass matrix can be obtained in our scheme.
We first discuss the bilinear model.
As can be seen in Table I, a bilinear model with low $\tan\beta$ produces
two heavy neutrino masses in the right range for the atmospheric and solar 
neutrino vacuum oscillation mass scale.  But, it appears to be impossible to
obtain bi-large mixing required in this case.   
In order to have bi-large mixing, one needs $U^\nu_{13} \ll 1$ 
which means $\xi_1 \ll \xi_{2,3}$ as the tree-level mass dominates \cite{jaja}.
This implies, in turn, $\mu_1 \ll \mu_{2,3}$ which gives rise to the
solar neutrino mixing angle $\sim \mu_1/\mu_{2,3}$,  leading  to a 
contradiction.
On the other hand, one can easily find a small mixing MSW solution to
solar neutrino problem \cite{bks} for a reasonably large $\tan\beta$.
A solution is as follows.
Let us take the input values with $\tan\beta=24$;
\begin{equation} \label{sol1}
 {\mu_1\over\mu}= 6\times10^{-6}\,,\quad
 {\mu_2\over\mu}=  {\mu_3\over\mu}= 1.5\times10^{-4}\,.
\end{equation}
This gives rise to the neutrino mass matrix in the eV  unit;
\begin{eqnarray} \label{matrix1}
&&-\pmatrix{ 5.7\times10^{-5} & 1.4\times10^{-3} & 1.3\times10^{-3} \cr
           1.4\times10^{-3} & 3.6\times10^{-2} & 3.2\times10^{-2} \cr
	  1.3\times10^{-3} & 3.2\times10^{-2} & 2.9\times10^{-2} \cr}
 +\pmatrix{ 2.2\times10^{-5} & 5.5\times10^{-4} & 2.3\times10^{-4} \cr
           5.5\times10^{-4} & 1.4\times10^{-2} & 5.8\times10^{-3} \cr
	  2.3\times10^{-4} & 5.8\times10^{-3} & 2.5\times10^{-3} \cr} 
	   \nonumber \\
&&~~~~~~~~~~~~~~~~~~~~~~~~~~~
= -\pmatrix{ 3.5\times10^{-5} & 8.7\times10^{-4} & 1.1\times10^{-3} \cr
           8.7\times10^{-4} & 2.2\times10^{-2} & 2.6\times10^{-2} \cr
           1.1\times10^{-3} & 2.6\times10^{-2} & 2.6\times10^{-2} \cr},
\end{eqnarray}
where the first matrix on the left-hand side is the tree-level mass 
and the second one is the one-loop mass.
This leads to the mass eigenvalues and oscillation amplitude;
\begin{eqnarray}
&& m_1=4.6\times10^{-19},\quad  
 m_2=2.3\times10^{-3},\quad
 m_3=   5.1\times10^{-2}       \nonumber  \\
&&\sin^22\theta_{\rm sol}=0.0035,\quad
  \sin^22\theta_{\rm atm}=0.99,\quad
  \sin^22\theta_{\rm chooz}=0.0029 \,.
\end{eqnarray}
In our notation, $\sin^22\theta_{\rm sol}$ is the $\nu_e$ disappearance 
oscillation amplitude corresponding to $\Delta m^2_{21}$,  and 
$\sin^22\theta_{\rm atm}$ ($\sin^22\theta_{\rm chooz}$)
is the $\nu_\mu\to \nu_\tau$ ($\nu_e$ disappearance) oscillation amplitudes 
corresponding to $\Delta m^2_{32,31}$, which are measured in the
solar neutrino \cite{sola}, and the atmospheric neutrino \cite{skam} 
(CHOOZ \cite{chooz}) experiments, respectively.
Let us note that such a solution can be obtained in a limited 
range of $\tan\beta=22-26$.  If one reduces the mediation scale, a larger
$\tan\beta$ is needed to achieve this kind of solution;  
e.g., $\tan\beta \approx 40$ for $\Lambda=50$ TeV.
Our investigation shows that the small mixing MSW solution to the solar
neutrino problem can be obtained for rather smaller $\tan\beta$ than
claimed in Ref.~\cite{hwang}.  The main reason for this discrepancy
is that the possibility of a precise cancellation for large $\tan\beta$ 
discussed above was not considered legitimately in a qualitative analysis
of Ref.~\cite{hwang}.  In fact, for $\tan\beta > 30$ (with $\Lambda=75$ TeV),
the loop corrections get weird due to a large third generation effect 
($h_b, h_\tau \sim 1$)  so that the input, $\mu_2 \approx \mu_3$, 
motivated by large atmospheric neutrino mixing, does not work at all.

\medskip

In trilinear models, we can have more freedom and thus find more variety
of solutions.  A remarkable result is that the bi-maximal mixing solution 
for the atmospheric and solar neutrino vacuum oscillations 
\cite{bimaximal} can be obtained in our framework.  For $\tan\beta=2$, 
a set of the input values to provide bi-maximal mixing is
\begin{eqnarray} \label{sol2}
&& \lambda_1= 2.3\times10^{-6}\,,\quad
 \lambda_2= 7.7\times10^{-6}\,,   \nonumber\\
&& \lambda'_1= -9.4\times10^{-7}\,,\quad
 \lambda'_2= \lambda'_3= -1.3\times10^{-5}\,.
\end{eqnarray}
A key feature to find such a  solution is that
one needs $\lambda'_1$ smaller than $\lambda'_{2,3}$
by about one order of magnitude to yield $U^\nu_{13} \ll 1$ and 
rather large $\lambda_1$ to increase the mass components in the
$\nu_e$ direction. The resultant mass matrix is given by
\begin{eqnarray} \label{matrix2}
&&-\pmatrix{ 1.3\times10^{-4} & 1.8\times10^{-3} & 1.7\times10^{-3} \cr
           1.8\times10^{-3} & 2.7\times10^{-2} & 2.5\times10^{-2} \cr
	  1.7\times10^{-3} & 2.5\times10^{-2} & 2.3\times10^{-2} \cr}
 +\pmatrix{ 6.7\times10^{-6} & 2.9\times10^{-5} & 1.6\times10^{-5} \cr
           2.9\times10^{-5} & 1.1\times10^{-4} & 5.0\times10^{-5} \cr
	  1.6\times10^{-5} & 5.0\times10^{-5} & -2.2\times10^{-6} \cr} 
	   \nonumber \\
&&~~~~~~~~~~~~~~~~~~~~~~~~~~~
= -\pmatrix{ 1.2\times10^{-4} & 1.8\times10^{-3} & 1.7\times10^{-3} \cr
           1.8\times10^{-3} & 2.7\times10^{-2} & 2.5\times10^{-2} \cr
           1.7\times10^{-3} & 2.5\times10^{-2} & 2.3\times10^{-2} \cr},
\end{eqnarray}
leading to the following mass eigenvalues and oscillation amplitudes;
\begin{eqnarray}
&& m_1=2.4\times10^{-6},\quad  
 m_2=8.1\times10^{-6},\quad
 m_3=5.0\times10^{-2}       \nonumber  \\
&&\sin^22\theta_{\rm sol}=0.97,\quad
  \sin^22\theta_{\rm atm}=0.99,\quad
  \sin^22\theta_{\rm chooz}=0.0098 \,.
\end{eqnarray}
In this case, we have $\Delta m^2_{21}=0.6\times10^{-10} {\rm eV}^2$.
Compared to the other solutions, this is rather sensitive to the
input values, but one does not need to rely on fine-tuning of the input
parameters.  

Our last example is to accommodate the small mixing MSW solar neutrino 
solution even for low $\tan\beta$.   It is the case where
$\lambda_i$ is larger than $\lambda'_i$.
Specifically, the input values with $\tan\beta=5$;
\begin{eqnarray}
&& \lambda_1=1.6\times10^{-6}\,,\quad
 \lambda_2=4.0\times10^{-5}\,  \nonumber\\
&& \lambda'_1=4.9\times10^{-7}\,,\quad
 \lambda'_2=\lambda'_3=1.2\times10^{-5}\,,
\end{eqnarray}
give rise to the mass matrix;
\begin{eqnarray} \label{matrix3}
&&-\pmatrix{ 4.0\times10^{-5} & 1.0\times10^{-3} & 8.5\times10^{-4} \cr
           1.0\times10^{-3} & 2.5\times10^{-2} & 2.1\times10^{-2} \cr
	  8.5\times10^{-4} & 2.1\times10^{-2} & 1.8\times10^{-2} \cr}
 +\pmatrix{ 4.4\times10^{-6} & 1.1\times10^{-4} & -7.5\times10^{-6} \cr
           1.1\times10^{-4} & 2.7\times10^{-3} & -1.9\times10^{-4} \cr
	  -7.5\times10^{-6} & -1.9\times10^{-4} & -1.4\times10^{-6} \cr} 
	   \nonumber \\
&&~~~~~~~~~~~~~~~~~~~~~~~~~~~
= -\pmatrix{ 3.6\times10^{-5} & 8.9\times10^{-4} & 8.5\times10^{-4} \cr
           8.9\times10^{-4} & 2.2\times10^{-2} & 2.1\times10^{-2} \cr
           8.5\times10^{-4} & 2.1\times10^{-2} & 1.8\times10^{-2} \cr}.
\end{eqnarray}
The corresponding  mass eigenvalues and oscillation amplitudes are
\begin{eqnarray}
&& m_1=1.5\times10^{-20},\quad  
 m_2=1.4\times10^{-3},\quad
 m_3=4.2\times10^{-2}       \nonumber  \\
&&\sin^22\theta_{\rm sol}=0.0029,\quad
  \sin^22\theta_{\rm atm}=0.99,\quad
  \sin^22\theta_{\rm chooz}=0.0035 \,.
\end{eqnarray}
In this solution, the loop mass matrix is dominated by 
the diagrams with $\lambda\lambda$ couplings and the other diagrams can
be neglected.
In all the realistic solutions, we find that 
the mass matrix elements coming from the diagrams with $h_\tau\lambda$ and  
$h_b\lambda'$ in $\Pi_D$, and with $gh_\tau$ and $h_\tau h_\tau$ in 
$\Pi_{D,n}$ are smaller than the others by more than one order of magnitude.
Furthermore, the $\Pi_D$ diagrams can be omitted as they give negligible
contributions to the resulting mass eigenvalues and mixing angles.
Among the $\Pi_n$ diagrams, one should not ignore  the 
diagrams with $g\lambda, h_\tau\lambda$ in the parameter region with
large $\tan\beta$, or relatively large $\lambda$ 
(as in the first two solutions),
as their contributions become comparable to those with $\lambda\lambda$ or 
$\lambda'\lambda'$ which usually give largest contributions.

\section {Conclusions}

We have performed a comprehensive analysis on the one-loop corrected
neutrino masses and mixing.   Our first concern was to determine
sneutrino vacuum expectation values at the one-loop level by employing 
one-loop effective scalar potential.  R-parity violation mixes
R-parity even and odd particles and thus sfermion and fermion mass
matrices are more complicated than in the R-parity conservation case.
However, under the condition that R-parity violating couplings are small
as implied by small neutrino masses, one can make an analytic  
diagonalization up to the quadratic order in R-parity violating parameters
to compute the complete one-loop contributions to sneutrino minimization 
condition.  The results are presented in Appendix A.
This one-loop determination of sneutrino VEVs  is essential 
to confront the R-parity violating models with experimental data.
It has been observed that the one-loop correction can give rise to a 
significant change from the tree-level calculation, in particular, 
when there occurs cancellation between tree-level contributions.  
Taking gauge-mediated supersymmetry breaking models, 
we have shown that this kind of cancellation
arises naturally for large $\tan\beta$ and the one-loop correction can 
change even the order of magnitude of the heaviest neutrino mass.

We have then considered one-loop renormalization of neutrino mass matrix
to determine mass eigenvalues and mixing angles for all three neutrinos.
Concerning the neutrino sector only, it turns out that the use of
the $Q$-independent effective mixing matrix is very useful to calculate
one-loop corrected neutrino masses and mixing.  In this procedure,
we compute 7x7 one-loop neutrino/neutralino mass matrix in the weak basis,
and make a $Q$-independent seesaw diagonalization to obtain
3x3 effective neutrino mass matrix.  From this effective mass matrix, one
can simply calculate mass eigenvalues and mixing matrix working at zero
momentum $p^2=0$.  It is then argued that this procedure simplifies
the calculation in the basis of $\mu_i=0$ if one wants to calculate
the leading contributions which are summarized in Appendix B.

Following the above procedures, we have investigated the properties of
neutrino mass matrix to see whether the atmospheric and solar neutrino
oscillations can be obtained in the framework of R-parity violating
theories with gauge mediated supersymmetry breaking.
In the bilinear model,  together with the atmospheric neutrino oscillation,
one can only obtain the small mixing MSW solution to the solar neutrino 
problem taking large $\tan\beta \gtrsim 20$.
For the case of the trilinear model, one can have more solutions.
With low $\tan\beta$, the bi-maximal mixing solution to
the atmospheric and solar neutrino problems can be realized.
Furthermore, varying 5 free parameters, it has been shown that
the small mixing MSW solution can be obtained even for small $\tan\beta$.
We would like to emphasize that all of these R-parity violating solutions 
to neutrino problems can be realized without any fine-tuning, apart from 
the smallness of R-parity violating parameters, $\mu_i, \lambda$ and 
$\lambda'$, which may be originated from a certain flavor symmetry 
explaining the quark and lepton mass hierarchies.

\appendix
\section{One-loop Contributions to Sneutrino VEVs}

We collect here one-loop contributions to sneutrino minimization 
condition (\ref{svev})
by calculating all the field-dependent particle masses.

{\it Stop contribution}:
The stop mass matrix including R-parity violating contribution
is given by
\begin{equation}
 {\cal M}^2_{\tilde{t}}= \pmatrix{
       m^2_{Q_3}+m_t^2+{1\over6}(4 M_W^2-M_Z^2)c_{2\beta} & 
       A_t v_2 + h_t(\mu v_1+\mu_i u_i) \cr
       A_t v_2 + h_t(\mu v_1+\mu_i u_i)  &
       m^2_{\tilde{t}_R}+m_t^2-{2\over3}( M_W^2-M_Z^2)c_{2\beta} \cr }
\end{equation}
where $h_t$ and $A_t$ are the  top quark Yukawa coupling  and trilinear 
soft parameter respectively,  and 
$M_W^2c_{2\beta}= g_w^2(v_1^2+u^2-v_2^2)$,
$M_Z^2c_{2\beta}= g_z^2(v_1^2+u^2-v_2^2)$.
Following the procedure described in Section II, we find the contributions
from the stop mass eigenstates $\tilde{t}_{1,2}$ as follows:
\begin{eqnarray}
\Sigma^{(1)}_{L_i}(\tilde{t}_j) &=& {6\over 64\pi^2} \left\{ 
  {4M^2_{Dt} \over 2m^2_{\tilde{t}_j} - m^2_{\tilde{t}_1}- m^2_{\tilde{t}_2} }
   {m_t \mu_i \over v^2s_{2\beta}/2} \right\} 
   m^2_{\tilde{t}_j}\!\left( {m^2_{\tilde{t}_j} \over Q^2} -1\right)
               \nonumber\\
\Sigma^{(2)}_{L_i}(\tilde{t}_j)&=& {6\over 64\pi^2}  \left\{
 {1\over2}{M_Z^2\over v^2} + { M^2_{Lt}-M^2_{Rt} \over 
    2m^2_{\tilde{t}_j} - m^2_{\tilde{t}_1}- m^2_{\tilde{t}_2} }
   { 8M^2_W-5M_Z^2 \over 6 v^2 } \right\}
   m^2_{\tilde{t}_j}\!\left({m^2_{\tilde{t}_j} \over Q^2} -1\right)
\end{eqnarray}
where $M^2_{Lt}= {\cal M}^2_{\tilde{t},11}$, 
$M^2_{Dt}={\cal M}^2_{\tilde{t},12}$ and  
$M^2_{Rt}={\cal M}^2_{\tilde{t},22}$ ignoring  small R-parity violating 
contributions.

{\it Sbottom contribution}:
The sbottom mass matrix is
\begin{equation}
 {\cal M}^2_{\tilde{b}}= \pmatrix{
       m^2_{Q_3}+m_b^2-{1\over6}(2 M_W^2+M_Z^2)c_{2\beta} & 
       -A_b v_2 - h_b\mu v_1+ A'_i u_i \cr
       -A_b v_2 - h_b\mu v_1+ A'_i u_i &
       m^2_{\tilde{b}_R}+m_b^2+{1\over3}( M_W^2-M_Z^2)c_{2\beta} \cr }
\end{equation}
The sbottom contributions are
\begin{eqnarray}
\Sigma^{(1)}_{L_i}(\tilde{b}_j) &=& {6\over 64\pi^2} \left\{ 
  2\lambda'_i{m_b\over v_1} + {4M^2_{Db} \over 2m^2_{\tilde{b}_j} - 
  m^2_{\tilde{b}_1}- m^2_{\tilde{b}_2} }{A'_i \over v_1} \right\}
  m^2_{\tilde{b}_j}\!\left( {m^2_{\tilde{b}_j} \over Q^2} -1\right) \\
\Sigma^{(2)}_{L_i}(\tilde{b}_j)&=& {6\over 64\pi^2}  \left\{
 -{1\over2}{M_Z^2\over v^2} + { M^2_{Lb}-M^2_{Rb} \over 
 2m^2_{\tilde{b}_j} - m^2_{\tilde{b}_1}- m^2_{\tilde{b}_2} }
 { -4M^2_W+M_Z^2 \over 6 v^2 }  \right\}
 m^2_{\tilde{t}_j}\!\left({m^2_{\tilde{t}_j} \over Q^2} -1\right)
                           \nonumber 
\end{eqnarray}
where $M^2_{Lb}$, $M^2_{Rb}$, and $M^2_{Db}$ are defined as in the stop case.
The contributions from the first two squark generations can be obtained
by obvious substitutions.  

{\it Charged slepton/Higgs contribution}:
Let us denote the charged slepton and Higgs contributions to $\Sigma$'s as
\begin{equation}
\Sigma^{(1,2)}_{L_j}(\phi) = {4\over 64\pi^2} 
 S^{(1,2)}_j(\phi)
 m^2_\phi\!\left(\ln{m^2_\phi\over Q^2}-1\right) \,,
\end{equation}
where $\phi$ stands for the mass eigenstates $\tilde{e}_{i1}, \tilde{e}_{i2}$
and $H^-$.
Here $S^{(1)}_j(\phi) \equiv  
 (\partial m^2_\phi/\partial u_j)/2v_1$  and 
$S^{(2)}_j(\phi) \equiv (\partial m^2_\phi/\partial u_j)/2u_j$  
can be calculated from the diagonalization of the mass matrix (
\ref{MlH},\ref{MlHs}).  In the matrix (\ref{MlHs}),  
the R-parity conserving mass components are 
\begin{eqnarray}
M^2_{Li}&=&m^2_{L_i} + m^{e2}_i - {1\over2}(2M^2_W-M^2_Z)  c_{2\beta},
      \nonumber\\
M^2_{Di} &=& -A^e_i v_1-h^e_iv_2\mu 
      \nonumber\\
M^2_{Ri} &=& m^2_{\tilde{e}_{iR}} + m^{e2}_i+(M^2_W-M^2_Z)   c_{2\beta}
\end{eqnarray}
apart from $u^2$ dependence in the gauge boson masses. The slepton mass
diagonalization matrix elements $c_{ei}$ and $s_{ei}$ define the
rotation angle as $\tan\theta_{ei}=s_{ei}/c_{ei}$ and useful quantities
which are free from the ambiguity in defining the angle $\theta_{ei}$ are
\begin{equation} \label{cisi}
 \sin{2\theta_{ei}}= {-2 M^2_{Di} \over 
         m^2_{\tilde{e}_{i1}} -m^2_{\tilde{e}_{i2}} }, \quad
 \cos{2\theta_{ei}}= {M^2_{Li}-M^2_{Ri} \over 
       m^2_{\tilde{e}_{i1}} -m^2_{\tilde{e}_{i2}} }
\end{equation}
A straightforward calculation leads to 
\begin{eqnarray}
&&S^{(1)}_j( \tilde{e}_{i1}, \tilde{e}_{i2} ) 
 =  {\delta_{ij} \over 2v_1^2 F_i}
 \left\{ 2m^e_i v_1\lambda_i F_i 
         -(m^2_{L_iH_1}+B_i t_\beta)c^2_\beta F_i    
	        \right.     \nonumber\\
&&\qquad  +(m^2_{L_iH_1}t_\beta-B_i)
    \left((M^2_{Ri}-m^2_{H^-})M^2_Wc_\beta^2s_{2\beta}  +
         M^4_{Di}{s_{2\beta}\over2}-m^e_i\mu M^2_{Di} \right) \nonumber\\
&&\qquad -(m^e_i\mu_i) \left(m^e_i\mu(M^2_{Li}-m^2_{H^-})/c^2_\beta
        -M^2_{Di}(M^2_{Li}-m^2_{H^-} +2M^2_Wc_\beta^2)t_{\beta} \right)  
	   \nonumber \\
&&\qquad \pm 
    {M^2_{Li}-M^2_{Ri} \over m^2_{\tilde{e}_{i1}}-m^2_{\tilde{e}_{i2}} } 
  \left[ -(m^2_{L_iH_1}+B_i t_\beta)c^2_\beta F_i
  +(m^e_i\mu_i) (m^e_i\mu)(M^2_{Li}-m^2_{H^-})/c^2_\beta
       \right. \nonumber \\
&&\qquad  \left.
   +(m^2_{L_iH_1}t_\beta-B_i) 
   \left(M^2_{Ri}-m^2_{H^-}) M^2_Wc_\beta^2s_{2\beta} 
          + M^4_{Di}{s_{2\beta}\over2} \right) \right]
         \nonumber\\
&&\qquad \pm {2M^2_{Di} \over m^2_{\tilde{e}_{i1}}-m^2_{\tilde{e}_{i2}} } 
  \left[ A_i v_1  F_i
          +m^e_i\mu_i t_\beta  F_i
                    \right.\nonumber\\
&&\qquad  -{1\over2}(m^e_i\mu_i)        
    \left((M^2_{Li}+M^2_{Ri}-2m^2_{H^-})(M^2_{Li}-m^2_{H^-}+2M^2_Wc^2_\beta)
      t_{\beta} -2m^e_i\mu M^2_{Di}/c^2_\beta \right)
               \nonumber\\
&&\qquad +{1\over2}(m^2_{L_iH_1}t_\beta-B_i)  \left.\left.
    \left(m^e_i\mu(M^2_{Li}+M^2_{Ri}-2m^2_{H^-})
     -M^2_{Di} (M^2_{Li}-m^2_{H^-}+2M^2_Wc^2_\beta)s_{2\beta} \right)
      \right] \right\}  \nonumber \\
&&S^{(1)}_i(H^-) = - {m^2_{L_iH_1}t_\beta-B_i \over v_1^2} 
 \left( (M^2_{Ri}-m^2_{H^-})M^2_Wc_\beta^2s_{2\beta}
   +M^4_{Di}{s_{2\beta}\over2}+M^2_{Di}m^e_i\mu   \right)
            \nonumber\\
&&\qquad +{m^e_i\mu_i \over v_1^2}
  \left( m^e_i\mu(M^2_{Li}-m^2_{H^-})/c^2_\beta
   -M^2_{Di}(M^2_{Li}-m^2_{H^-} +2M^2_Wc_\beta^2)t_{\beta}  \right)
\end{eqnarray}
 where  $F_i \equiv (m^2_{\tilde{e}_{i1}}-m^2_{H^-}) 
                   (m^2_{\tilde{e}_{i2}}-m^2_{H^-}) $. \\
Note that the contributions for the two light sleptons are
reduced to the following simple form taking $m^e_{1,2}=0$:
\begin{eqnarray}
S^{(1)}_i( \tilde{e}_{iL} ) &=&  -{m^2_{L_iH_1}+B_it_\beta \over v^2}
  +{m^2_{L_iH_1}t_\beta-B_i \over v^2}
      {M^2_Ws_{2\beta}   \over (M^2_{Li}-m^2_{H^-})  }  \nonumber\\
S^{(1)}_i( \tilde{e}_{iR} ) &=& 0 
\end{eqnarray} 
Similarly, one finds
\begin{eqnarray}
&&S^{(2)}_j(\tilde{e}_{i1}, \tilde{e}_{i2}) =
 -{1\over4}{M^2_Z \over v^2}  
 \pm { M^2_{Li}-M^2_{Ri} \over m^2_{\tilde{e}_{i1}}-m^2_{\tilde{e}_{i2}} }
 \left({3\over4} {M_Z^2\over v^2} - {M_W^2\over v^2} \right)
	       \nonumber\\
&&~~~~~+{1\over2}{\delta_{ij} \over v_1^2 F_i}
 \left\{  \left( m^2_{H^-}s^2_\beta -M^2_{Li}- M_W^2c_\beta^2 
          \right)  F_i \right.
    + m^{e2}_i\mu^2(M^2_{Li}-m^2_{H^-})/c^2_\beta 
          \nonumber\\
&&\qquad+(M^2_{Ri}-m^2_{H^-})(M^2_{Li}-m^2_{H^-}+2M^2_Wc^2_\beta)^2s^2_\beta
    -2m^e_i\mu t_\beta M^2_{Di}(M^2_{Li}-m^2_{H^-}+2M^2_Wc^2_\beta) 
            \nonumber\\
&&\qquad \pm {M^2_{Li}-M^2_{Ri} \over 
     m^2_{\tilde{e}_{i1}}-m^2_{\tilde{e}_{i2}} }\left[
   \left( m^2_{H^-}s^2_\beta-M^2_{Li} -M^2_W c_\beta^2 \right)F_i 
    -  m^{e2}_i\mu^2(M^2_{Li}-m^2_{H^-})/c^2_\beta 
      \right.  \nonumber\\
&&\qquad \qquad~~~~~~~~~~~~~ \left.
 +  (M^2_{Ri}-m^2_{H^-})(M^2_{Li}-m^2_{H^-}+2M^2_Wc^2_\beta)^2 s^2_\beta
         \right]    \nonumber\\
&&\qquad \pm { 2M^2_{Di} \over m^2_{\tilde{e}_{i1}}-m^2_{\tilde{e}_{i2}} }
     \left[ \left(-2A_i v_1- M^2_{Di} \right)  F_i - 
      M^2_{Di} \left( m^{e2}_i\mu^2/c^2_\beta
              +(M^2_{Li}-m^2_{H^-}+2M^2_Wc^2_\beta)^2 s^2_\beta \right) 
           \right.\nonumber\\
&&\qquad \left.\left. ~~~~~~~~~~~~~~~~~~~
     + m^e_i\mu t_\beta(M^2_{Li}+M^2_{Ri}-2m^2_{H^-})  
        (M^2_{Li}-m^2_{H^-}+2M^2_Wc^2_\beta) 
	     \right]\right\}  
	            \nonumber \\
&&S^{(2)}_i(H^-) = {1\over v^2} \left\{
  \left( M^2_W(1+2s_\beta^2)  - {1\over2}M^2_Z c_{2\beta}
  + (M^2_{Li}-m^2_{H^-})t^2_\beta \right) F_i \right.\nonumber\\
&&~~~~~~~~~~~~~~
 \qquad-(M^2_{Ri}-m^2_{H^-})(M^2_{Li}-m^2_{H^-}+2M^2_Wc^2_\beta)^2s^2_\beta
         +m^{e2}_i\mu^2(M^2_{Li}-m^2_{H^-})/c^2_\beta 
                  \nonumber\\
&&~~~~~~~~~~~~~~\qquad \left.
   +2 m^e_i\mu t_\beta M^2_{Di} (M^2_{Li}-m^2_{H^-}+2M^2_Wc^2_\beta) \right\}
\end{eqnarray}

{\it CP-even sneutrino/Higgs contribution}:
As in the above case, we write
\begin{equation}
\Sigma^{(1,2)}_{L_i}(\phi) = {2\over 64\pi^2} 
 S^{(1,2)}_i(\phi) 
 m^2_\phi\!\left(\ln{m^2_\phi\over Q^2}-1\right) \,,
\end{equation}
where $\phi$ runs for ${\rm Re}(\tilde{\nu})$, $h^0$ and $H^0$ and   
$S^{(1),(2)}_i(\phi)$ are calculated from the matrix (\ref{MRH}) as follows:
\begin{eqnarray}
S^{(1)}_j(\tilde{\nu}_i) &= & -{\delta_{ij} \over F_i}
      {M_Z^2\over v^2} \left(
       m^2_{L_i H_1}(m^2_{A^0}c_{2\beta}-m^2_{\tilde{\nu}_i})
       +B_it_\beta(m^2_{A^0}c_{2\beta}+m^2_{\tilde{\nu}_i}) \right) 
              \nonumber\\
S^{(1)}_i(h^0,H^0) &= & {1\over2}{M_Z^2 \over  v^2}{1\over F_i} \left\{
   \left(  m^2_{L_i H_1}(m^2_{A^0}c_{2\beta}-m^2_{\tilde{\nu}_i})
          +B_it_\beta(m^2_{A^0}c_{2\beta}+m^2_{\tilde{\nu}_i}) \right) 
	            \right. \nonumber\\
&& \mp c_{2\alpha}
\left(m^2_{L_i H_1} (m^2_{A^0}c^2_{\beta}+M^2_Zs^2_\beta-m^2_{\tilde{\nu}_i})
   +B_it_\beta (m^2_{A^0}s^2_{\beta}+M^2_Zc^2_\beta-m^2_{\tilde{\nu}_i}) 
              \right) \nonumber\\
&& \mp \left.  {s_{2\alpha}\over2}
\left( (m^2_{L_i H_1}t_\beta +B_i)(m^2_{A^0}+M^2_Z)c_{2\beta}+
       2(m^2_{L_i H_1}t_\beta -B_i)m^2_{\tilde{\nu}_i} \right)
             \right\}     \nonumber\\
S^{(2)}_j(\tilde{\nu}_i)&=& {1\over2}{M_Z^2\over v^2} 
  \left[ 1 + \delta_{ij} \left( 1 + 
    {M^2_Z(m^2_{\tilde{\nu}_i}-m^2_{A^0}c^2_{2\beta}) \over F_i} 
         \right) \right] \nonumber\\
S^{(2)}_i(h^0,H^0)&=& {1\over2}{M_Z^2 \over  v^2} \left\{
  { M^2_Z(m^2_{A^0}c^2_{2\beta}-m^2_{\tilde{\nu}_i}) \over F_i }
 \mp c_{2\alpha}
 \left[ 1+ {M^2_Z(m^2_{A^0}-m^2_{\tilde{\nu}_i})c_{2\beta}\over F_i} \right] 
         \right.\nonumber\\
&&\left.
  \mp s_{2\alpha} s_{2\beta}
     {M^2_Z m^2_{\tilde{\nu}_i} \over F_i}  \right\}  
\end{eqnarray}      
where $ F_i  \equiv m^4_{\tilde{\nu}_i}
      -2m^2_{\tilde{\nu}_i}(m^2_{h^0}+ m^2_{H^0}) + m^2_{h^0} m^2_{H^0}$,
and $ m^2_{\tilde{\nu}_i}=m^2_{L_i}+{1\over2}M_Z^2c_{2\beta}$. \\
Recall that 
\begin{equation}
c_{2\alpha}=-c_{2\beta} {m^2_{A^0}-M^2_Z \over m^2_{H^0}- m^2_{h^0} },\quad
s_{2\alpha}=-s_{2\beta}{m^2_{A^0}+M^2_Z \over m^2_{H^0}- m^2_{h^0} }.
\end{equation}

{\it CP-odd sneutrino/Higgs contribution}:
Again, one can calculate from the matrix (\ref{MIH}), 
\begin{equation}
\Sigma^{(1,2)}_{L_j}(\phi) = {2\over 64\pi^2} 
 S^{(1,2)}_i(\phi)
 m^2_\phi\!\left(\ln{m^2_\phi\over Q^2}-1\right) \,,
\end{equation}
where $\phi$ runs for ${\rm Im}(\tilde{\nu})$, $A^0$.
Here, $S^{(1),(2)}_i(\phi)$ are given by 
\begin{eqnarray}
&&S^{(1)}_j( \tilde{\nu}^I_i)= 
      -\delta_{ij}{m^2_{L_iH_1}+B_it_\beta \over v^2},\quad 
S^{(1)}_i( A^0)=  0   \nonumber\\
&&S^{(2)}_j( \tilde{\nu}^I_i)= {1\over2}{M^2_Z \over v^2}
      -\delta_{ij}{m^2_{\tilde{\nu}_i} \over v^2},\quad ~~~~
S^{(2)}_i( A^0)= -{1\over2} {M^2_Z \over v^2} c_{2\beta} 
\end{eqnarray}

{\it Gauge boson contribution}:

\begin{eqnarray}
\Sigma^{(1)}_{L_i}(\phi)&=& 0 \nonumber\\
\Sigma^{(2)}_{L_i}(\phi)&=& {12\over 64\pi^2} 
 {M^4_W \over v^2}\!\left(\ln{M^2_W\over Q^2}-1\right) 
                           +{6\over 64\pi^2} 
 {M^4_Z \over v^2}\!\left(\ln{M^2_Z\over Q^2}-1\right)  
\end{eqnarray}

{\it Bottom quark contribution}:

\begin{eqnarray}
\Sigma^{(1)}_{L_i}(\phi)&=& -{24\over 64\pi^2} {m_b \over v_1} \lambda_{i}
 {m^2_b \over v^2}\!\left(\ln{m^2_b\over Q^2}-1\right)  \nonumber\\
\Sigma^{(2)}_{L_i}(\phi)&=& 0
\end{eqnarray}

{\it Neutrino/neutralino contribution}:
To diagonalize the 7x7 neutrino/neutralino mass matrix, we use
the seesaw formula to obtain the rotation matrix:
\begin{equation} \label{UN}
 {\cal U}^N = \pmatrix{ 1 & \Theta_N \cr -\Theta_N^\dagger & 1 \cr}
              \pmatrix{  U^\nu & 0 \cr 0 &  N \cr}
\end{equation}
where the seesaw diagonalization matrix $\Theta_N$ separating  out
the neutrino and neutralino parts is given by $\Theta_N=M_D M^{-1}_N$, and
$U^\nu$ is the neutrino mixing matrix which is determined from the full 
neutrino mass matrices including the tree and one-loop contributions.  
The matrix $N$ is the usual neutralino diagonalization matrix whose 
analytic expression can be found, e.g., in the paper by Barger
{\it et al.} in Ref.~\cite{hvev}.
We present a simpler form as follows:
\begin{eqnarray}  \label{Ns}
&& {N_{2i} \over N_{1i}} = -{1\over t_W}
             {m_{\chi^0_i}-M_1 \over m_{\chi^0_i}-M_2 } \nonumber \\
&& {N_{3i} \over N_{1i}} = -{1\over s_W}
             {(m_{\chi^0_i}-M_1)(m_{\chi^0_i}c_\beta-\mu s_\beta)
	     \over M_Z (m_{\chi^0_i}- \mu s_{2\beta})  }\\
&& {N_{3i} \over N_{1i}} = -{1\over s_W}
             {(m_{\chi^0_i}-M_1)(m_{\chi^0_i}s_\beta-\mu c_\beta)
	     \over M_Z (m_{\chi^0_i}- \mu s_{2\beta}) } \nonumber 
\end{eqnarray}
and the $N_{ij}$ satisfies the unitarity relation,
$N_{1i}^2+N_{2i}^2+N_{3i}^2+N_{4i}^2=1$.
Recall that one has to keep R-parity violating contributions to calculate
$\Sigma$'s which we do not show explicitly in, e.g., (\ref{Ns}).

The sneutrino VEVs contributions are
\begin{equation}
\Sigma^{(1,2)}_{L_i}(\psi)= -{8\over 64\pi^2} 
  S^{(1,2)}_i(\psi) 
 m^2_\psi \!\left(\ln{m^2_\psi\over Q^2}-1\right) 
\end{equation}
where $\psi$ runs for four neutralinos $\chi^0_i$, and 
$S^{(1,2)}(\psi)$ are given by
\begin{eqnarray}
S^{(1)}_i(\psi)&=& \mu_i {M^2_Z \over v^2} m_\psi \left[
     -m^2_\psi t_\beta + m_\psi( M_{\tilde{\gamma}}t_\beta+\mu)
     -\mu M_{\tilde{\gamma}} \right] /D(\psi)
                           \nonumber\\
S^{(2)}_i(\psi)&=& {M^2_Z \over v^2} m_\psi \left[ -m^3_\psi 
  + m^2_\psi  M_{\tilde{\gamma}} 
  + m_\psi \mu^2- \mu^2 M_{\tilde{\gamma}} \right]/D(\psi)  \,,
\end{eqnarray}
where
\begin{eqnarray}
 D(\psi) &=& 5m^4_\psi-4m^3_\psi(M_1+M_2)+3m^2_\psi(M_1M_2-\mu^2-M^2_Z) 
                      \nonumber\\
 &&+2m_\psi[\mu^2(M_1+M_2)+M^2_Z(2\mu s_{2\beta}+ M_{\tilde{\gamma}})] 
   +\det(M_N) \,.  \nonumber
\end{eqnarray}
Note that $S^{(1)}_i=0$ in the basis with $\mu_i=0$.

{\it Charged lepton/chargino contribution}:
Diagonalization matrices ${\cal U}^R, {\cal U}^L$ bring respectively
the weak eigenstates $(e^c_i, \tilde{W}^+, \tilde{H_2}^+)$ and 
$(e_i, \tilde{W}^-, \tilde{H_1}^-)$ to the mass eigenstates 
$(e^\pm_i, \chi^\pm_{1,2})$ and are given by
\begin{eqnarray} \label{ULR}
&& {\cal U}^R =
 \pmatrix{ 1 & r_1^i c_r + r_2^i s_r  & -r_1^i s_r + r_2^i c_r \cr
           -r_1^i & c_r & -s_r \cr
	   -r_2^i & s_r & c_r \cr} \nonumber\\
&& {\cal U}^L =
 \pmatrix{ 1 & l_1^i c_l + l_2^i s_l  & -l_1^i s_l + l_2^i c_l \cr
           -l_1^i & c_l & -s_l \cr
	   -l_2^i & s_l & c_l \cr} 
\end{eqnarray}
where $r_{1,1}$ and $l_{1,2}$ are given by
\begin{eqnarray} \label{rls}
 r_1^i= ({u_i \over v_1} -{\mu_i\over \mu})
      { \sqrt{2}m^e_i \mu M_W (\mu-M_2 t_\beta) c_\beta \over 
           (M_2\mu+M^2_Ws_{2\beta})^2 },\quad  &&
 r_2^i= ({u_i \over v_1} -{\mu_i\over \mu})
      { m^e_i \mu (M_2^2+2M_W^2 c^2_\beta)  \over 
           (M_2\mu+M^2_Ws_{2\beta})^2 }            \nonumber\\
 l_1^i= ({u_i \over v_1} -{\mu_i\over \mu})
      { \sqrt{2} \mu M_W  c_\beta \over M_2\mu+M^2_Ws_{2\beta} },\quad  &&
 l_2^i= ({u_i \over v_1} -{\mu_i\over \mu})
      { M_W^2 s_{2\beta} \over M_2\mu+M^2_Ws_{2\beta} }  +{\mu_i\over \mu}
\end{eqnarray}
and the useful combination of the usual chargino mixing angles  are 
given by
\begin{eqnarray}
&& c_lc_r = { M_2 m_{\chi_1^-}+ \mu m_{\chi_2^-} \over
               m^2_{\chi_1^-} - m^2_{\chi_2^-} },\quad
 c_ls_r = { \sqrt{2} M_W (s_\beta m_{\chi_1^-}+ c_\beta m_{\chi_2^-})\over
               m^2_{\chi_1^-} - m^2_{\chi_2^-} }  \nonumber \\
&& s_ls_r = { \mu m_{\chi_1^-}+ M_2 m_{\chi_2^-} \over
               m^2_{\chi_1^-} - m^2_{\chi_2^-} }  , \quad
 s_lc_r = { \sqrt{2} M_W (c_\beta m_{\chi_1^-}+ s_\beta m_{\chi_2^-})\over
               m^2_{\chi_1^-} - m^2_{\chi_2^-} } \,.
\end{eqnarray}
Neglecting the contributions from the two light charged leptons
by taking $m_{e,\mu}=0$, we get
\begin{equation}
\Sigma^{(1,2)}_{L_i}(\psi)= -{8\over 64\pi^2} S^{(1,2)}_i(\psi) 
 m^2_\psi\!\left(\ln{m^2_\psi\over Q^2}-1\right) 
\end{equation}
where $\psi$ runs for $\tau$ and the charginos  $\chi^-_{1,2}$ and 
$S^{(1,2)}(\psi)$ are given by
\begin{eqnarray}
&&S^{(1)}_i(\psi) = \mu_i {m^2_\psi \over v^2} \left[
   2M^2_W(\mu-M_2t_\beta)(m^2_\psi-m^2_\tau) 
    +\delta_{i3}m^2_\tau\mu\left(2M^2_W+(m^2_\psi-M^2_2) /c^2_\beta 
     \right) \right] /D(\psi) \nonumber\\
&&~~~~~~~~~~~ -\lambda_i m_\tau{m^2_\psi \over v_1} \left[
     (m^2_{\psi}-m^2_{\chi^+_1})(m^2_{\psi}-m^2_{\chi^+_2})\right]/D(\psi)
                 \\
&&S^{(2)}_i(\psi) = 2 {m^2_\psi \over v^2} \left[
 m^4_\psi M^2_W - m^2_\psi M^2_W(\mu^2+2M^2_W s^2_\beta+m^2_\tau)
 +m^2_\tau M^2_W(\mu^2+2M^2_W s^2_\beta)  \right. 
                      \nonumber\\
&&~~~~
 \left. + \delta_{i3} m^2_\tau \left\{ m^4_\psi/2c^2_\beta + m^2_\psi
  (M^2_W(3-t^2_\beta) -M^2_2/c_\beta^2)   
 -2 M^2_W(M_2\mu t_\beta-3M^2_Ws^2_\beta-\mu^2)\right\}\right]/D(\psi)  
                   \nonumber
\end{eqnarray}
where
$$ D(\psi)= 5m^6_\psi -4m^4_\psi(m^2_{\chi^-_1}+ m^2_{\chi^-_2} +m^2_\tau)  
+3 m^2_\psi [ m^2_\tau (m^2_{\chi^-_1}+m^2_{\chi^-_2}) 
         +  m^2_{\chi^-_1} m^2_{\chi^-_2} ]
      -2m^2_\tau m^2_{\chi^-_1} m^2_{\chi^-_2} \,. $$

\section{One-loop Corrections to Neutrino Masses}

Here we summarize the calculation of one-loop self-energy diagrams in the
weak basis.  A key ingredient in this procedure is to express 
the propagators inside loops in terms of the mass basis propagators which 
can be done by the use of the diagonalization matrices 
(\ref{UC},\ref{US},\ref{UP},\ref{UN},\ref{ULR}).
Schematically, one can write the transformation from the weak basis 
propagators to the mass basis ones as follows:
\newcommand{\p}{{\cal P}}
\newcommand{\U}{{\cal U}}
\begin{equation}
 \langle \alpha \beta \rangle
  = \sum_{\psi} \p_{\alpha\beta}^{\psi}   
 \langle \psi \psi \rangle
\end{equation}
where $\alpha$ and $\beta$ represent weak basis fields and 
$\psi$ mass eigenstates.
For neutrino/neutralino, charged lepton/chargino, 
sneutrino/Higgs (scalar or pseudoscalar), and charged slepton/Higgs
propagators, one finds
\begin{eqnarray} \label{cP}
\p_{\rho^0_1\rho^0_2}^{\psi^0} = 
     \U^N_{\rho^0_1\psi^0}\U^N_{\rho^0_2\psi^0}, &\quad& 
\p_{\rho^-_1\rho^+_2}^{\psi^\pm} = 
     \U^L_{\rho^-_1\psi^-}\U^R_{\rho^+_2\psi^+} 
	 \nonumber\\
\p_{\varphi_1^0\varphi_2^0}^{\phi^0} = 
    \pm{1\over2} \U^{S,P}_{\varphi_1^0 \phi^0}\U^{S,P}_{\varphi_2^0\phi^0},
                  &\quad&
\p_{\varphi_1^0\varphi_2^{0*}}^{\phi^0} = 
    {1\over2} \U^{S,P}_{\varphi_1^0 \phi^0}\U^{S,P}_{\varphi_2^{0*}\phi^0} 
           \nonumber\\
\p_{\varphi_1^-\varphi_2^{+}}^{\phi^-} = 
     \U^{C}_{\varphi_1^- \phi^-}\U^{C}_{\varphi_2^+\phi^+}  &\quad&
\end{eqnarray}
where $\rho$'s and $\psi$'s represent the neutrino/neutralino or
charged lepton/chargino fields in the weak and mass basis, respectively;
$\varphi^0$ is a  complex sneutrino or Higgs field and 
$\phi^0$ is a scalar or pseudoscalar mass eigenstates;
$\varphi^-$ and $\phi^-$ represent the charged slepton/Higgs fields
in the weak and mass basis, respectively.

We first present $\Pi_n$ which gives most significant contributions
to the  one-loop neutrino self-energy.  
\begin{eqnarray}  \label{Pin}
 \Pi_{\nu_i\nu_j} &=& \left\{
   -{g^2 \over32\pi^2} \sum_{\psi^0,\phi^0} 
      \p_{\tilde{Z} \tilde{Z}}^{\psi^0} 
     \p_{\tilde{\nu} \tilde{\nu} }^{\phi^0}  
      m_{\psi^0}B_0(p^2,m^2_{\psi^0},m^2_{\phi^0}) 
       \right. \nonumber\\
&&   -{gh_i \over16\pi^2} 
      \sum_{\psi^-,\phi^-}
      \p_{\tilde{W}^- e^c_i}^{\psi^-} \p_{\tilde{e}_j H_1^+}^{\phi^-}
      m_{\psi^-}B_0(p^2,m^2_{\psi^-},m^2_{\phi^-}) 
               \nonumber\\
&&  -{g \lambda_i\over 16\pi^2} \delta_{j3} \sum_{\psi^-,\phi^-}
      \p_{\tilde{W}^- \tau^c}^{\psi^-}
     (\p_{\tilde{\tau}_L \tilde{\tau}_L^*}^{\phi^-} -
      \p_{\tilde{e}_{iL} \tilde{e}_{iL}^*}^{\phi^-})
      m_{\psi^-}B_0(p^2,m^2_{\psi^-},m^2_{\phi^-}) \nonumber\\
&&  -{  h_i h_j\over 16\pi^2}  \sum_{\psi^-,\phi^-}
      \p_{\tilde{H}^-_1 e^c_i}^{\psi^-}\p_{H_1^- \tilde{e}_{jR}^*}^{\phi^-} 
      m_{\psi^-}B_0(p^2,m^2_{\psi^-},m^2_{\phi^-}) \nonumber\\
&&  -{h_\tau \lambda_j\over 16\pi^2} \delta_{i3}  \sum_{\psi^-,\phi^-}
      (\p_{\tau \tau^c}^{\psi^-}\p_{H_1^- \tilde{\tau}_R^*}^{\phi^-}
      +\p_{\tilde{H_1}^-\tau^c}^{\psi^-}
       \p_{\tilde{\tau}_L \tilde{\tau}_R^*}^{\phi^-})
      m_{\psi^-}B_0(p^2,m^2_{\psi^-},m^2_{\phi^-}) \nonumber\\
&&  -{ \lambda_i \lambda_j \over16\pi^2} \sum_{\psi^-,\phi^-}
      \p_{\tau \tau^c}^{\psi^-}\p_{\tilde{\tau}_L \tilde{\tau}_R^*}^{\phi^-}
      m_{\psi^-}B_0(p^2,m^2_{\psi^-},m^2_{\phi^-}) \nonumber\\
&& \left. -{ \lambda'_i \lambda'_j \over16\pi^2} \sum_{k}
      \p_{\tilde{b}_L \tilde{b}_R^*}^{\tilde{b}_k} 
      m_{b}B_0(p^2,m^2_{b},m^2_{\tilde{b}_k}) \right\}
      +(i\leftrightarrow j)
\end{eqnarray}

The contributions to $\Pi_D$ are
\begin{eqnarray}
\Pi_{\nu_i \tilde{B}}&=&
   -{g^2\over16\pi^2}{t_W\over2} \sum_{\psi^0,\phi^0}
   [-\p_{\tilde{W}_3\tilde{H}_1^0}^{\psi^0}
     \p_{\tilde{\nu}_i H_1^0}^{\phi^0}
   +\p_{\tilde{W}_3\tilde{H}_2^0}^{\psi^0}
    \p_{\tilde{\nu}_i H_2^0}^{\phi^0}]
   m^2_{\psi^0} B_0(p^2,m^2_{\psi^0},m^2_{\phi^0}) \nonumber\\
&& -{g^2\over16\pi^2}{t_W\over\sqrt{2}}\sum_{\psi^-,\phi^-}
   [2\p_{\tilde{W}^- e^c_i}^{\psi^-}
     \p_{\tilde{e}_{iL}\tilde{e}_{iR}^*}^{\phi^-}
   +\p_{\tilde{W}^-\tilde{H}_2^+}^{\psi^-}
    \p_{\tilde{e}_{iL} H_2^+}^{\phi^-}]
   m^2_{\psi^-} B_0(p^2,m^2_{\psi^-},m^2_{\phi^-}) \nonumber\\
&& -{g h_i\over16\pi^2}{t_W\over\sqrt{2}}\sum_{\psi^-,\phi^-}
   [-\p_{e_i e^c_i}^{\psi^-}
     \p_{\tilde{e}_{iL} H_1^+}^{\phi^-}
   -\p_{\tilde{H}_1^- e^c_i}^{\psi^-}
    \p_{H_1^- H_1^+}^{\phi^-}            \nonumber\\
&&~~~~~~~~~~~~~~~~~~
   +2\p_{\tilde{H_1}^- e^c_i}^{\psi^-}
     \p_{\tilde{e}_{iR} \tilde{e}_{iR}^*}^{\phi^-}
   +\p_{\tilde{H}_1^- \tilde{H}_2^+}^{\psi^-}
    \p_{ \tilde{e}_{iR} H_2^+}^{\phi^-} ]
   m^2_{\psi^-} B_0(p^2,m^2_{\psi^-},m^2_{\phi^-}) \nonumber\\
&& -{g\lambda_i \over16\pi^2}{t_W\over\sqrt{2}}\sum_{\psi^-,\phi^-}
   \p_{\tau \tau^c}^{\psi^-}
   [2\p_{\tilde{\tau}_{iR}\tilde{\tau}_{iR}^*}^{\phi^-}
    -\p_{\tilde{\tau}_{iL} \tilde{\tau}_{iL}^*}^{\phi^-}]
   m^2_{\psi^-} B_0(p^2,m^2_{\psi^-},m^2_{\phi^-}) \nonumber\\
&& -{g\lambda'_i \over16\pi^2}{t_W\over3\sqrt{2}}\sum_{k}
   [2\p_{\tilde{b}_{iR}\tilde{b}_{iR}^*}^{\tilde{b}_k}
    +\p_{\tilde{b}_{iL} \tilde{b}_{iL}^*}^{\tilde{b}_k}]
   m^2_b B_0(p^2,m^2_b,m^2_{\tilde{b}_k}) 
       \\
\Pi_{\nu_i \tilde{W}_3}&=&
   -{g^2\over16\pi^2}{1\over2} \sum_{\psi^0,\phi^0}
   [\p_{\tilde{W}_3\tilde{H}_1^0}^{\psi^0}
    \p_{\tilde{\nu}_i H_1^0}^{\phi^0}
   -\p_{\tilde{W}_3\tilde{H}_2^0}^{\psi^0}
    \p_{\tilde{\nu}_i H_2^0}^{\phi^0}]
   m^2_{\psi^0} B_0(p^2,m^2_{\psi^0},m^2_{\phi^0}) \nonumber\\
&& -{g^2\over16\pi^2}{1\over\sqrt{2}}\sum_{\psi^-,\phi^-}
   [\p_{\tilde{W}^- \tilde{H}_2^+}^{\psi^-}
     \p_{\tilde{e}_{iL} H_2^+}^{\phi^-}]
   m^2_{\psi^-} B_0(p^2,m^2_{\psi^-},m^2_{\phi^-}) \nonumber\\
&& -{g h_i\over16\pi^2}{1\over\sqrt{2}}\sum_{\psi^-,\phi^-}
   [-\p_{e_i e^c_i}^{\psi^-}
     \p_{\tilde{e}_{iR} H_1^+}^{\phi^-}
   -\p_{\tilde{H}_1^- e^c_i}^{\psi^-}
    \p_{H_1^- H_1^+}^{\phi^-}    
   +\p_{\tilde{H}_1^- \tilde{H}_2^+}^{\psi^-}
    \p_{ \tilde{e}_{iR} H_2^+}^{\phi^-} ]
   m^2_{\psi^-} B_0(p^2,m^2_{\psi^-},m^2_{\phi^-}) \nonumber\\
&& -{g\lambda_i \over16\pi^2}{1\over\sqrt{2}}\sum_{\psi^-,\phi^-}
   [-\p_{\tau \tau^c}^{\psi^-} 
    \p_{\tilde{\tau}_{iL} \tilde{\tau}_{iL}^*}^{\phi^-}]
   m^2_{\psi^-} B_0(p^2,m^2_{\psi^-},m^2_{\phi^-}) \nonumber\\
&& -{g\lambda'_i \over16\pi^2}{1\over\sqrt{2}}\sum_{k}
   [- \p_{\tilde{b}_{iL} \tilde{b}_{iL}^*}^{\tilde{b}_k}]
   m^2_{b} B_0(p^2,m^2_{b},m^2_{\tilde{b}_k}) 
          \\
\Pi_{\nu_i \tilde{H}^0_1}&=&
   -{g^2\over16\pi^2}{1\over2} \sum_{\psi^0,\phi^0}
   [-t_W\p_{\tilde{W}_3\tilde{B}}^{\psi^0}
    \p_{\tilde{\nu}_i H_1^0}^{\phi^0}
   +\p_{\tilde{W}_3\tilde{W}_3}^{\psi^0}
    \p_{\tilde{\nu}_i H_1^0}^{\phi^0}]
   m^2_{\psi^0} B_0(p^2,m^2_{\psi^0},m^2_{\phi^0}) \nonumber\\
&& -{g h_i\over16\pi^2}\sum_{\psi^-,\phi^-}
   [-\p_{\tilde{W}^- e^c_i}^{\psi^-}
     \p_{\tilde{e}_{iL} \tilde{e}_{iL}^*}^{\phi^-}
    +\p_{\tilde{W}^- e^c_i}^{\psi^-}
     \p_{H_1^-H_1^+}^{\phi^-}]
   m^2_{\psi^-} B_0(p^2,m^2_{\psi^-},m^2_{\phi^-}) \nonumber\\
&&-{h_i^2\over16\pi^2} \sum_{\psi^-,\phi^-}
   [-\p_{e_i e^c_i}^{\psi^-}
     \p_{\tilde{e}_{iR} H_1^+}^{\phi^-}
    -\p_{\tilde{H}^-_1 e^c_i}^{\psi^-}
     \p_{\tilde{e}_{iL} \tilde{e}^*_{iR}}^{\psi^-}]
   m^2_{\psi^-} B_0(p^2,m^2_{\psi^-},m^2_{\phi^-}) \nonumber\\
&& -{h_\tau\lambda_i \over16\pi^2}\sum_{\phi^-}
   [-2 \p_{\tau_i \tau^c_i}^{\psi^-}
    \p_{\tilde{\tau}_{iL} \tilde{\tau}_{iR}^*}^{\phi^-}]
   m^2_{\psi^-} B_0(p^2,m^2_{\psi^-},m^2_{\phi^-}) \nonumber\\
&& -{h_b\lambda'_i \over16\pi^2}\sum_{k}
   [-2 \p_{\tilde{b}_{iL} \tilde{b}_{iR}^*}^{\tilde{b}_k}]
   m^2_b B_0(p^2,m^2_b,m^2_{\tilde{b}_k}) 
                 \\
\Pi_{\nu_i \tilde{H}^0_2}&=&
   -{g^2\over16\pi^2}{1\over2} \sum_{\psi^0,\phi^0}
   [t_W\p_{\tilde{W}_3\tilde{B}}^{\psi^0}
    \p_{\tilde{\nu}_i H_2^0}^{\phi^0}
   +\p_{\tilde{W}_3\tilde{W}_3}^{\psi^0}
    \p_{\tilde{\nu}_i H_2^0}^{\phi^0}]
   m^2_{\psi^0} B_0(p^2,m^2_{\psi^0},m^2_{\phi^0}) \nonumber\\
&& -{g^2  \over16\pi^2}\sum_{\psi^-,\phi^-}
   [\p_{\tilde{W}^- \tilde{W}^+}^{\psi^-}
     \p_{\tilde{e}_{iL} H_2^+}^{\phi^-}]
   m^2_{\psi^-} B_0(p^2,m^2_{\psi^-},m^2_{\phi^-})  \nonumber\\
&& -{g h_i\over16\pi^2}\sum_{\psi^-,\phi^-}
   [\p_{\tilde{H}_1^- \tilde{W}^+}^{\psi^-}
     \p_{\tilde{e}_{iR} H_2^+}^{\phi^-}]
   m^2_{\psi^-} B_0(p^2,m^2_{\psi^-},m^2_{\phi^-}) 
\end{eqnarray}
Note that $\p_{e_i e^c_i}^{\psi^-}$ is equal to $1$ for $\psi^-=e_i$ and 
vanishes for all others, and that 
 $\p_{\tilde{b}_L \tilde{b}_R^*}^{\tilde{b}_{1,2}} = \mp c_b s_b $,
 $\p_{\tilde{b}_L \tilde{b}_L^*}^{\tilde{b}_{1,2}} = c_b^2, s_b^2 $, and 
 $\p_{\tilde{b}_R \tilde{b}_R^*}^{\tilde{b}_{1,2}} = s_b^2, c_b^2 $
where $c_b$ and $s_b$ define the sbottom diagonalization angle; 
$\tan\theta_b= s_b/c_b$ as in Eq.~(\ref{cisi}).  
The function $B_0$ is defined by
\begin{equation}
 B_0(p^2,m_1^2,m_2^2)=-{m_2^2 \over m_1^2-m_2^2} \ln {m_1^2 \over m_2^2}
        -\ln{m_1^2 \over Q^2} +1 
\quad{\rm with}\quad  p^2 \to 0 .
\end{equation}

\section{RGE in the basis with no bilinear terms}

Renormalization group equations for the lepton number violating 
parameters in the basis where the bilinear terms $\mu_i L_iH_1$ in
the superpotential are rotated away have been calculated in the second
paper of Ref.~\cite{cw}.  Here we collect the results 
with a few minor corrections.
\begin{eqnarray}
16\pi^2 {d \lambda'_i \over d t} &=& 
  \lambda'_i ( \delta_{i3} h_\tau^2 + h_t^2 + 3 h_b^2 
   - {7\over9} g^{\prime2} - 3 g_2^2 -{16\over3}g_3^2) \\
16\pi^2 {d \lambda_i \over d t} &=& 
   \lambda_i ( 3 h_\tau^2  - 3 g^{\prime2} - 3 g_2^2) \\
16\pi^2 {d A'_i \over d t} &=& 
   A'_i ( \delta_{i3} h_\tau^2 + h_t^2 + 9 h_b^2 
   - {7\over9} g^{\prime2} - 3 g_2^2 -{16\over3}g_3^2) + A_i(2h_b h_\tau) \\
   &+ & 2 \lambda'_i ( \delta_{i3} h_\tau A_\tau + h_t A_t + 3 h_b A_b
   + {7\over9} g^{\prime2} M_1+ 3 g_2^2M_2 +{16\over3}g_3^2M_3)
\nonumber\\
16\pi^2 {d A_i \over d t} &=& 
   A_i ( 5 h_\tau^2  - 3g^{\prime2} - 3 g_2^2) + A'_i(6h_b h_\tau) 
   +\lambda_i ( 6 h_\tau A_\tau + 6 g^{\prime2} M_1+ 6 g_2^2M_2)  \\
16\pi^2 {d m_{L_i H_1}^2 \over d t} &=& 
   m_{L_i H_1}^2 (  \delta_{i3} h_\tau^2 + h_\tau^2 + 3 h_b^2 ) 
   -6A'_i A_b - 2 A_i A_\tau  \\
   &-&6\lambda'_i h_b (m_{L_i}^2+m_{Q_3}^2+m_{D_3}^2) 
   -2\lambda_i h_\tau (m_{L_i}^2+m_{L_3}^2+m_{E_3}^2) \nonumber \\
16\pi^2 {d B_i \over d t} &=& B_i ( \delta_{i3} h_\tau^2 
      +3 h_t^2-g^{\prime2}-3g_2^2) -\mu (6 A'_i h_b+2 A_i h_\tau)
\end{eqnarray}
Note that these equations are valid when the R-parity violating 
couplings are small enough so that the next leading corrections are
negligible.

\begin{table}
\caption{R-parity violating parameters $m^2_{L_iH_1}, B_i$,
and and one-loop corrections $\Sigma_{L_i}$ in the GeV unit.  
Each collum corresponds to the lepton generation, $i=1,2,3$.  
$\xi_i$ are one-loop improved sneutrino VEVs and $\xi^0_i$ are 
tree values in the unit of $v_1$.  The last five rows show 
typical sizes of some components of tree and loop mass neutrino 
matrices and their eigenvalues in the eV unit.
}
\medskip

\begin{tabular}{c|rrr||rrr}
  &  & $\tan\beta=3$ & & & $\tan\beta=30$ & \\ \hline
$m^2_{L_iH_1}$       &  
  $4.00\times10^{-2}$ &  $4.00\times10^{-2}$ &  $3.93\times10^{-2}$ &
  $3.67$      &   $3.67$     &   $3.59$   \\
$B_i\tan\beta$       &   
  $-3.95\times10^{-3}$ &  $ -3.95\times10^{-3}$ &  $ -3.88\times10^{-3}$ &
   $-3.60$     &      $-3.60$   &    $-3.52$  \\ \hline
\hfill$\tilde{q}~~$&
  $4.93\times10^{-3}$  &  $4.93\times10^{-3}$     &  $4.93\times10^{-3}$ &
  $5.02\times10^{-1}$  &    $5.02\times10^{-1}$   &  $5.00\times10^{-1}$ \\
\hfill$\phi^-$  &
  $1.39\times10^{-3}$    & $1.39\times10^{-3}$    &  $1.45\times10^{-3}$   &
   $-1.73\times10^{-2}$  &  $-1.73\times10^{-2}$  &  $-3.89\times10^{-2}$  \\
\hfill$S^0$  &
  $-5.64\times10^{-4}$   &    $-5.64\times10^{-4}$   & $-5.02\times10^{-4}$  &
  $-5.03\times10^{-2}$   &   $-5.03\times10^{-2}$    & $ -4.61\times10^{-2}$  \\
\hfill$P^0$  &
  $7.21\times10^{-4}$   &  $7.21\times10^{-4}$   &   $7.55\times10^{-4}$  &
  $1.48\times10^{-3}$   &  $1.48\times10^{-3}$   &   $1.39\times10^{-3}$  \\ 
$\Sigma^{(1)}_{L_i}$       &
  $6.47\times10^{-3}$  &     $6.47\times10^{-3}$   &   $6.63\times10^{-3}$  &
  $4.36\times10^{-1}$  &     $4.36\times10^{-1}$   &   $4.17\times10^{-1}$   \\  \hline
$m^2_{\tilde{\nu}_i}$  &
  $57280$   &   $57280$   &   $62860$   &
  $56440$   &   $56440$   &   $61340$   \\
\hfill$\tilde{q}~~$  &
  $-4.84$   &   $-4.84$   &    $-4.84$   &
  $-6.22$   &   $-6.22$   &    $-6.22$     \\
\hfill$\phi^-$  &
  $2507$   &   $2507$   &   $2867$   &
  $2382$   &   $2382$   &   $2178$    \\
\hfill$S^0$  &
  $-356.3$  &  $-356.3$  &   $-360.2$  &
  $-329.8$  &  $-329.8$  &   $-331.9$   \\
\hfill$P^0$  &
  $739.6$   &  $739.6$   &   $935.6$  &
  $759.3$   &  $759.3$   &   $928.9$  \\ 
\hfill$\psi^0$  &
  $-712.6$   &   $-712.6$  &  $-712.6$  &
  $-486.3$   &   $-486.3$  &  $-486.3$    \\
\hfill$\psi^-$  &
  $782.8$    &   $782.8$   &   $787.3$    &
  $745.0$    &   $745.0$   &   $1048$       \\
\hfill$Z/W$     &
  $-258.8$   &   $-258.8$   &   $-258.8$   &   
  $-258.8$   &   $-258.8$   &   $-258.8$     \\  
$\Sigma^{(2)}_{L_i}$  &
  $2696$   &   $2696$   &   $3253$   &
  $2805$   &   $2805$   &   $3072$       \\  \hline
$\xi_i$              &       
  $-7.09\times10^{-7}$  &  $-7.09\times10^{-7}$  &  $ -6.35\times10^{-7}$   &
  $-8.61\times10^{-6}$  &  $-8.61\times10^{-6}$ &  $ -7.50\times10^{-6}$  \\
$\xi^0_i$            &     
  $-6.30\times10^{-7}$  &  $-6.30\times10^{-7}$  &  $-5.63\times10^{-7}$  &
  $-1.32\times10^{-6}$  &  $-1.32\times10^{-6}$ &   $-1.08\times10^{-6}$ \\  
               \hline
tree mass &
  &   $2\times10^{-3}$  &  &
  &   $3\times10^{-3}$    \\
$\Pi_D:\eta~~$ &
  &   $10^{-5}~~~$      &    &
  &   $10^{-4}~~~$          \\
$\Pi_n:h\lambda$  &   
  &   $10^{-8}~~~$      &   &
  &   $10^{-2}~~~$        \\
$\Pi_n:\lambda\lambda$ &
  &   $10^{-5}~~~$      &  &
  &   $10^{-2}~~~$        \\
eigenvalues   & 
  0, & $1\times10^{-5}$, &  $7\times10^{-3}$ &
  0, & $7\times10^{-3}$, &  $2\times10^{-2}$ \\
\end{tabular}
\end{table}
\end{document}